\documentclass{article}

\usepackage{microtype}
\usepackage{graphicx}
\usepackage{subfigure}
\usepackage{booktabs} 

\usepackage{hyperref}

\usepackage[accepted]{icml2024}

\usepackage{amsmath}
\usepackage{amssymb}
\usepackage{mathtools}
\usepackage{amsthm}

\usepackage[capitalize,noabbrev]{cleveref}

\theoremstyle{plain}
\newtheorem{theorem}{Theorem}[section]

\theoremstyle{definition}

\newtheorem{assumption}[theorem]{Assumption}
\theoremstyle{remark}

\usepackage[textsize=tiny]{todonotes}

\icmltitlerunning{The Statistical Significance of the Inclusion of Graph Neural Networks in the  Financial Time Series Forecasting Problem}

\begin{document}

\twocolumn[
\icmltitle{The Statistical Significance of the Inclusion of Graph Neural Networks\\ in the Financial Time Series Forecasting Problem}



\icmlsetsymbol{equal}{*}

\begin{icmlauthorlist}
\icmlauthor{Marco Gregnanin}{equal,imt,ku}
\icmlauthor{Johannes De Smedt}{equal,ku}
\icmlauthor{Giorgio Gnecco}{equal,imt}
\icmlauthor{Maurizio Parton}{equal,uni}
\end{icmlauthorlist}

\icmlaffiliation{imt}{Laboratory for the Analysis of compleX Economic Systems (AXES), IMT School for Advanced Studies Lucca, Italy}
\icmlaffiliation{ku}{Research Centre for Information Systems Engineering (LIRIS), KU Leuven, Belgium}
\icmlaffiliation{uni}{Department of Economic Studies, University of Chieti–Pescara, Italy}

\icmlcorrespondingauthor{Marco Gregnanin}{marco.gregnanin@imtlucca.it}

\icmlkeywords{Graph Neural Networks, Time Series Forecasting}

\vskip 0.3in
]



\printAffiliationsAndNotice{\icmlEqualContribution} 

\begin{abstract}
Forecasting univariate time series in the financial market is a challenging endeavor. While numerous statistical and machine learning models have been introduced to address this challenge, they typically concentrate solely on analyzing temporal patterns within the time series data.
In this research, we study the statistical significance of the inclusion of geometric patterns in enhancing forecasting accuracy within the context of time series analysis. We introduce the Time-Geometric model, a combination 
of models designed to exploit both geometric and temporal patterns.  
The contribution of this research lies in advancing the domain of univariate time series prediction,
as demonstrated through extensive empirical evaluations. Our findings underscore that leveraging geometric patterns, captured through Graph Neural Networks, yields statistically significant improvements in forecasting accuracy.
\end{abstract}

\section{Introduction}
Univariate time series forecasting remains a challenging task due to inherent noise within the data, with no universally accepted method for noise removal and pattern identification \cite{Bishop2016}. Financial time series, in particular, pose additional challenges due to their stochastic nature \cite{Tsay2005, lamberton2011introduction}.\\
Various methods, spanning statistical, financial mathematical, and deep learning models, address the forecasting problem for financial time series. Statistical models like AutoRegressive Integrated Moving Average (ARIMA) \cite{Box2015} assume stationarity in stock prices. Financial mathematical models leverage Geometric Brownian motion \cite{Black1973pricing} and jump processes \cite{tankov2003financial}. Deep learning models, such as Recurrent Neural Networks (RNNs) \cite{Goodfellow2016book}, Transformers \cite{vaswani2017attention}, and Temporal Convolutional Networks (TCNs) \cite{lea2016temporal}, have shown success in exploiting temporal patterns within univariate time series \cite{elman1990finding, hua2019deep}. However, these models predominantly focus on temporal patterns, and the performance improvements demonstrated often lack an analysis of statistical significance, made according to established criteria \cite{demvsar2006statistical}.\\
This research investigates 
the statistical significance of the inclusion of geometric patterns in enhancing forecasting accuracy for various types of time series models, including RNNs, Transformers, and TCNs. By representing univariate time series as graphs, we transition from Euclidean to non-Euclidean spaces, enabling the identification of complex patterns through graph structures, as indicated by the degree distribution. The ability to discern fractal, random, or periodic time series, particularly relevant in financial markets, is facilitated through the graph-based approach \cite{lacasa2008time, mandelbrot2013fractals}. Graph Neural Networks (GNNs) \cite{scarselli2008graph} prove effective in capturing geometric patterns by iteratively aggregating information from neighboring nodes in a graph \cite{hamilton2017inductive}. To address this, we propose a model combination named the ``Time-Geometric'' model, which combines baseline time series neural network models with dynamic GNNs. We leverage the visibility graph method (which is traditionally applied in complex network analysis) to derive a graph representation of a univariate time series.
Our contributions include:
\begin{itemize}
    \item[1.] The introduction of the Time-Geometric model, which combines time series neural network models with dynamic GNNs;
    \item[2.] Its extensive evaluation across 90 financial univariate time series, which reveals performance improvements when considering geometric patterns;
    \item[3.] A statistical analysis of the results, based on both a pairwise and a multiple comparison approach, which highlights metric-dependent statistical differences among models.   
\end{itemize}
The paper is structured as follows: Section \ref{section_2} reviews relevant GNN models for time series and statistical comparisons among different algorithms; Section \ref{section_3} defines the research problem; Section \ref{section_4} outlines the proposed model; Section \ref{section_5} analyzes our proposed model compared to baseline models; Section \ref{section_6} studies the statistical significance of the inclusion of geometric patterns, and finally, Section \ref{section_7} concludes the paper.

\section{Related Works}
\label{section_2}
In this section, we provide a comprehensive review of relevant GNN models applied to time series and discuss pertinent research on multiple comparisons among algorithms. For an in-depth exploration of the forecasting problem for univariate time series, incorporating both statistical and deep learning models, please refer to Appendix \ref{appendix:related work}.\\
GNNs have demonstrated success across diverse domains, including traffic forecasting, network analysis, recommendation systems, and analysis of biological and chemical data \cite{Wu2021survey}. Noteworthy GNN models include the Spatio-Temporal Graph Convolutional Network (STGCN) \cite{Yu2018stgcn} for traffic prediction, the Spectral Temporal GNN (StemGNN) \cite{cao2020spectral} tested on real-world time series datasets (e.g., traffic forecasting, COVID-19), the Temporal Graph Convolutional Network (T-GCN) \cite{Zhao2020tgcn} for traffic prediction, Graph Wavelet Neural Networks (GWNNs) \cite{xu2018graphwavelet} for classification tasks, and Graph Recurrent Neural Networks (GRNNs) \cite{ruiz2020gatedrnn} for earthquake epicenter, traffic, and epidemic tracking prediction. Specifically in the financial domain, models like multi-modality GNN (MAGNN) \cite{Cheng2018magnn} analyzed two Chinese stock exchanges, and Graph WaveNet with Self-Attention \cite{Zhao2023wavenetatt} implemented the GWNN model to predict crude oil prices. Temporal and Heterogeneous GNN (THGNN) \cite{Xiang2022thgnn} focused on predicting the United States and Chinese stock market behaviors. However, these models are primarily designed for multivariate time series, deriving graph representations from correlation matrices, attention mechanisms, or geographical features. Contrary to this, in \citet{lazcano2023combined}, the authors derived the graph representation of the crude oil prices using the visibility graph method to predict them.\\
Furthermore, in machine learning research, time-series models are often compared with various baseline models without a rigorous statistical analysis of the differences in results. Exceptions include \citet{barrera2022rainfall} comparing machine learning models for predicting rainfall, \citet{Shih2019ComparisonOT} comparing statistical and deep learning models for blood supply. A deep statistical analysis was made in \citet{Parmezan2019} comparing different time series models over 95 datasets, and revealing no statistical differences in the results. Lastly, \citet{demvsar2006statistical} analyzed the statistical significance of several classifier models proposed at the International Conference on Machine Learning from 1999 to 2003.\\
To summarize, GNN models are generally applied to multivariate time series forecasting problems without undergoing statistical tests on the obtained results. Additionally, classical univariate time series models rely solely on the autoregressive nature of the time series, neglecting geometric patterns.

\section{Problem Formulation}
\label{section_3}
This research aims to demonstrate the statistical significance of the inclusion of geometric patterns in enhancing the predictive capacity of baseline neural network models for univariate financial time series forecasting.\\
Let $\mathbf{S}$ represent a collection of $F$ univariate time series, each comprising realizations over $T$ discrete time steps and denoted as $S^{i}=\{S^{i}_1,\dots, ,S^{i}_T\}$, $i \in \{1,\ldots,F\}$. The initial step involves defining a mapping function, $f(\cdot)$, capable of predicting the future $q$ values of a univariate target time series, denoted as $S^{*}$ and chosen from $\mathbf{S}$, based on the graph-based representation $G$ of the target time series and the collection of $F^{'} \leq F$ univariate time series correlated with $S^{*}$. Consequently, the problem is formulated as learning the function $f$ in such a way that 
\begin{equation}
    S^{*}_{t+1:t+q} = f(G_t, X_t) \,.
    \label{eq:f_function_definition}
\end{equation}
Here, $G_t$ represents the graph representation of the target time series $S^{*}$, obtained by considering the set $S^{*}_t = \{S^{*}_{t-m+1}, \dots, S^{*}_{t}\}$ of its past $m$ observations up to time $t$, and $X_t$ denotes an associated feature matrix, constructed by considering the set $S^{i}_t=\{S^{i}_{t-m+1}, \dots, S^{i}_{t}\}$, $i \in \{1, \dots, F^{'}\}$ of the last $m$ observations from each univariate time series $S^{i}$ from $\mathbf{S}$ that is correlated with $S^{*}$, 
which are represented in a tabular format. Additionally, a baseline function $g(\cdot)$ is defined as the mapping function predicting the future $q$ values of $S^{*}$ based solely on $X_t$:
\begin{equation}
    S^{*}_{t+1:t+q}=g(X_t)\,.
    \label{eq:g_function_definition}
\end{equation}
The subsequent step involves evaluating the performance of $K$ learning algorithms, obtained from $f(\cdot)$ and $g(\cdot)$ over $N$ different datasets. Let $v_{u}^{j}$ represent the value assumed by an evaluation metric of the $j$-th algorithm, computed on the $u$-th dataset.  
The aim is to assess whether the 
evaluation metric differences among the algorithms, both in pairwise and multiple comparisons, are statistically significant. To achieve this, two key assumptions are considered:
\begin{assumption}[Evaluation metrics comparison]
    A decrease 
    in the value of the 
    evaluation metric from $v_{u}^{1}$ to $v_{u}^{2}$ implies that algorithm $j=2$ performs better than algorithm $j=1$.
    \label{assumption_1}
\end{assumption}
\begin{assumption}[Reliability]
The evaluation metrics $v_{u}^{j}$ are reliable for all $K$ algorithms over the $N$ datasets, i.e., for each dataset, they are computed starting from the same random sample for all the algorithms. 
\label{assumption_2}
\end{assumption}
In the following, the goal is to compare, using statistical hypothesis testing, algorithms respectively based/not based on the use of geometric patterns for 
univariate time series forecasting. These employ functions, respectively, of the form (\ref{eq:f_function_definition}) and (\ref{eq:g_function_definition}).

\section{Proposed Time-Geometric Framework}
\label{section_4}

\begin{figure}[ht]
\vskip 0.1in
\begin{center}
\centerline{\includegraphics[width=1\columnwidth]{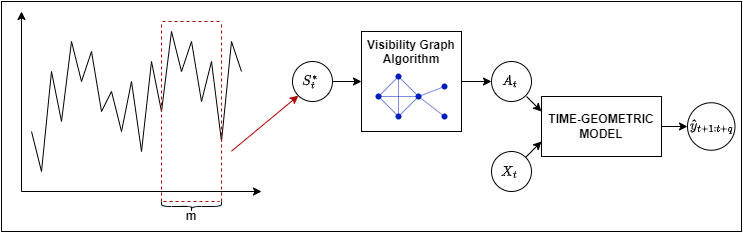}}
\caption{General idea of the proposed Time-Geometric Model.}
\label{fig:model}
\end{center}
\vskip -0.1in
\end{figure}

This section introduces the Time-Geometric model, a novel approach that effectively extracts and exploits both temporal and geometric patterns for univariate time series forecasting. Figure \ref{fig:model} illustrates the general architecture of the model. 

Initially, we begin with the set $S^{*}_t$ of the last $m$ observations of our target time series $S^{*}$. Subsequently, we employ the visibility graph algorithm to derive a graph-based representation of the time series, encoded using the adjacency matrix $A_t$\footnote{Definition provided in Appendix \ref{appendix:preliminaries_network}.}. Finally, the Time-Geometric model receives the adjacency matrix $A_t$ and the feature matrix $X_t$ as inputs. Thus, the Time-Geometric model corresponds to the mapping function defined in Equation (\ref{eq:f_function_definition}).

\subsection{Visibility Graph Algorithm}
\begin{figure}[ht]
\vskip 0.1in
\begin{center}
\centerline{\includegraphics[width=1\columnwidth]{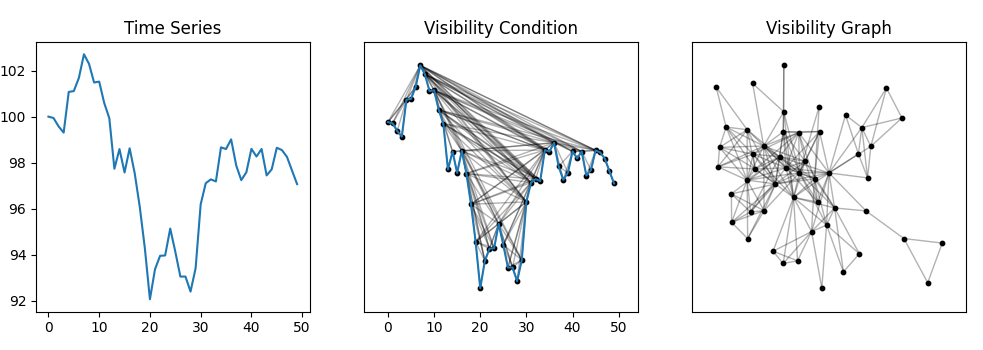}}
\caption{The Visibility Graph algorithm was applied to a time series generated from a Geometric Brownian Motion with the initial value equal to $100$ and with mean and variance equal to $0.05$ and $0.5$, respectively. The computation of the algorithm was conducted using the ``time series to visibility graphs'' (ts2vg) Python package \cite{Bergillos2020visibility}.}
\label{fig:visibility_graph}
\end{center}
\vskip -0.1in
\end{figure}

We opt for the visibility graph algorithm \cite{lacasa2008time} to construct the graph-based representation of the target time series $S^{*}$. This algorithm is chosen due to its capability to capture non-linear dependencies and patterns within time series data while preserving their structural properties. Depending on the nature of the time series — periodic, random, or fractal — we derive a regular graph, a random graph, or a small-world graph, respectively\footnote{Definitions provided in Appendix \ref{appendix:preliminaries_network}.} \cite{lacasa2008time}. The resulting graph representation is connected and invariant to affine transformations of the time series \cite{stephen2015visibility}.
Consider the target univariate time series $S^{*}=\{s^{*}_1, s^{*}_2, \dots, s^{*}_T\}$ associated with the set of discrete time steps $\mathcal{T}=\{1, 2, \dots, T\}$. The visibility graph algorithm represents the values of $S^{*}$ as vertical bins, forming edges whenever there is a clear line of sight between two data points. This implies that an edge is formed between each pair of bars if the top of one bar is visible from the other.
Each data point becomes a node in the graph, with edges representing visibility connections. The visibility condition, as defined in \citet{lacasa2008time}, states that two data points $(s^{*}_i,i)$ and $(s^{*}_j,j)$ are connected if any other data point $(s^{*}_k,k)$ satisfies: $s^{*}_k<s^{*}_j + (s^{*}_j-s^{*}_i)\frac{j - k}{j - i}$. This condition allows us to derive an undirected graph by considering all possible combinations of values for $i\in\{1,\dots,T\}$. Alternatively, we can obtain a directed graph by enforcing the ``left-to-right'' condition, comparing $(s^{*}_j,j)$ only with data points where $k > i$, i.e., $\forall(s^{*}_j,j)$ with $j=i+1, \dots, T$. The choice between a directed or undirected graph is an optimization hyperparameter. In Figure \ref{fig:visibility_graph}, we illustrate the steps of the visibility graph algorithm in obtaining the graph representation. The first plot displays the time series path, the middle plot illustrates the visibility condition and the link formation, and the final plot depicts the resulting undirected graph. The visibility graph can be computed for any length of the time series. However, in our model, we compute the graph representation considering the last $m$ observations. Therefore, the associated adjacency matrix $A_t$ has dimensions $m \times m$, with each node corresponding to a time observation.

\subsection{Time-Geometric Model}

\begin{figure}[ht]
\vskip 0.1in
\begin{center}
\centerline{\includegraphics[width=\columnwidth]{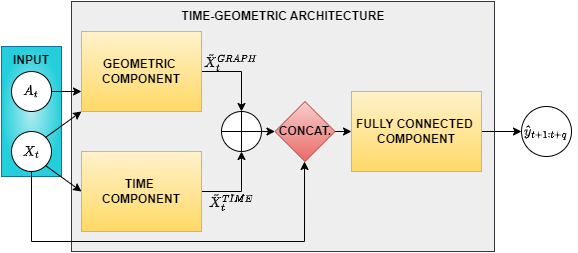}}
\caption{Time-Geometric Model Architecture. The baseline model differs in the absence of both the input $A_t$ and the Geometric Component.}
\label{fig:time_geometric}
\end{center}
\vskip -0.1in
\end{figure}

The Time-Geometric model receives input in the form of the adjacency matrix $A_t \in \mathbb{R}^{m \times m}$ and the feature matrix $X_t \in \mathbb{R}^{m \times F^{'}}$, where $m$ represents the number of past observations considered, and $F^{'}$ denotes the number of features. The model produces an output vector comprising $q$ elements, denoted as $\hat{y}_{t+1:t+q}$.\\
As illustrated in Figure \ref{fig:time_geometric}, the Time-Geometric model is comprised of three distinct components: the Time Component, the Geometric Component, and the Fully Connected Component. 

\subsubsection*{Time Component}
The Time Component takes the feature matrix $X_t \in \mathbb{R}^{m \times F^{'}}$ as input and produces the output $\Tilde{X}^{TIME}_t \in \mathbb{R}^{m \times F^{'}}$. Its objective is to analyze and leverage temporal patterns within the data. Algorithm \ref{alg:time_component} outlines the operations of this component.

\begin{algorithm}
   \caption{Time Component}
   \label{alg:time_component}
\begin{algorithmic}
   \STATE {\bfseries Input:} feature matrix $X_t$
   \STATE {\bfseries Output:} temporal pattern $\Tilde{X}^{TIME}_t$
   \STATE $R^{(0)}_t = X_t$ \qquad 
   \FOR{$l=1$ {\bfseries to} $L$}
   \STATE $R^{(l)}_t = \text{TIME}^{(l)}(R^{(l-1)}_t)$ 
   \ENDFOR
   \STATE $\Tilde{X}^{TIME}_t = \text{FC}(R^{(L)}_t)$ 
\end{algorithmic}
\end{algorithm}

The process involves initializing a hidden layer, denoted as $R_{t}^{(0)}$, with the feature matrix. Subsequently, based on the number of layers $L$, the $l$-th hidden layer of a neural network model for time series - $R^{(l)}_t = \text{TIME}^{(l)}(R^{(l-1)}_t)$ - is applied. Here, $\text{TIME}^{(l)}$ represents the considered neural network model, akin to the mapping function $g(\cdot)$ defined in Equation (\ref{eq:g_function_definition}). Specifically, for the Time Component analysis, the following neural network models are considered: Long Short-Term Memory (LSTM) \cite{hochreiter1997long}, Gated Recurrent Unit (GRU) \cite{chung2014empirical}, RNNs \cite{Goodfellow2016book}, Bidirectional RNNs (BiRNNs) \cite{schuster1997bidirectional}, Bidirectional LSTM (BiLSTM) \cite{graves2005bidirectional}, Bidirectional GRU (BiGRU) \cite{xiong2016dynamic}, Transformers \cite{vaswani2017attention}, and TCNs \cite{lea2016temporal}. Finally, to obtain the output of the Time Component, a fully connected layer denoted as $\text{FC}(\cdot)$, is applied, defined by:
\begin{equation}
    \Tilde{X}^{TIME}_t  = \phi \left( W_t R^{(L)}_t + b_t \right) \,,\label{eq:fully_connected_layer}
\end{equation}
where $W_t$ is the weight matrix, $b$ is a bias vector, and $\phi (\cdot)$ denotes the activation function, a hyperparameter to be determined.
 
\subsubsection*{Geometric Component}
The Geometric Component takes as input the adjacency matrix $A_t \in \mathbb{R}^{m \times m}$, derived using the visibility graph on the set $S^{*}_t$ of the last $m$ observations of the target time series $S^{*}$, and the feature matrix $X_t \in \mathbb{R}^{m \times F^{'}}$. The output of this block represents the geometric patterns and is denoted as $\Tilde{X}^{GRAPH}_t \in \mathbb{R}^{m \times F^{'}}$. Algorithm \ref{alg:geometric_component} outlines the operations of this component.

\begin{algorithm}[ht]
   \caption{Geometric Component}
   \label{alg:geometric_component}
\begin{algorithmic}
   \STATE {\bfseries Input:} adjacency matrix $A_t$, feature matrix $X_t$
   \STATE {\bfseries Output:} geometric pattern $\Tilde{X}^{GRAPH}_t$
   \STATE $H^{(0)}_t = X_t$ 
   \FOR{$l=1$ {\bfseries to} $L^{'}$}
   \STATE $H^{(l)}_t = \text{GNN}^{(l)}(H^{(l-1)}_t)$ 
   \ENDFOR
   \STATE $\Tilde{X}^{GRAPH}_t = \text{LSTM}(H^{(L^{'})}_t)$ 
\end{algorithmic}
\end{algorithm}

The process begins by initializing the hidden layer, denoted as $H^{(0)}_t$, with the feature matrix. Depending on the number of layers $L^{'}$, the $l$-th layer of a GNN is applied, denoted as $H^{(l)}_t = \text{GNN}^{(l)}(H^{(l-1)}_t)$. This can be defined using the message-passing paradigm:
\begin{align}
    M^{(l)}_{u \rightarrow v, t} &= \text{MSG}^{(l)}\left(H^{(l-1)}_{u,t}, H^{(l-1)}_{v,t}\right) \,,\nonumber \\
    H^{(l)}_{v,t} &= \text{AGG}^{(l)}\left( M^{(l)}_{u \rightarrow v, t} | u \in N(v), H^{(l-1)}_{v,t}\right) \,,\label{eq:gnn_equation}
\end{align}
where $\text{MSG}^{(l)}(\cdot)$ represents the message function bringing the message embedding at the $l$-th layer, $N(v)$ represents the neighborhood of node $v$, and $\text{AGG}^{(l)}(\cdot)$ is the aggregation function that combines the messages received from neighboring nodes. The role of the message function is to facilitate the exchange of information among nodes and their neighbors \cite{hamilton2017inductive}. The notation ``$l$-th layer'' denotes information originating from nodes that are at a distance of $l$-hops away.  Note that changing the definition of the message and aggregation function, as defined in Equation (\ref{eq:gnn_equation}), leads to different types of GNN models. In this research, we decide to consider only the Graph Convolutional Networks (GCNs) \cite{kipf2016semi}, and the $l$-th layer can be defined as:
\begin{equation*}
    H^{(l)}_t = \rho\left(\Tilde{D}_{t}^{-1/2}\Tilde{A}_{t} \Tilde{D}_{t}^{-1/2}H^{(l-1)}_t\Theta\right)\,,
\end{equation*}
where $\rho(\cdot)$ is the activation function, $\Theta$ is the weight matrix, $\Tilde{A}_{t} = A_t + I$, $I$ is the identity matrix, $\Tilde{D}_{t}$ is the degree matrix computed on $\Tilde{A}_{t}$. Before obtaining the output of the geometric component, we pass $H^{(L^{'})}_t$ through an LSTM layer to process the geometric patterns. 

\subsubsection*{Fully Connected Component}
The inputs of the Fully Connected Component are the temporal patterns $\Tilde{X}^{TIME}_t \in \mathbb{R}^{m \times F^{'}}$, obtained from the Time Component, the geometric patterns $\Tilde{X}^{GRAPH}_t \in \mathbb{R}^{m \times F^{'}}$, computed by the Geometric Component, and the feature matrix $X_t \in \mathbb{R}^{m \times F^{'}}$. Algorithm \ref{alg:FullyConnected_component} outlines the operations of this component.

\begin{algorithm}[ht]
   \caption{Fully Connected Component}
   \label{alg:FullyConnected_component}
\begin{algorithmic}
   \STATE {\bfseries Input:} temporal pattern $\Tilde{X}^{TIME}_t$, geometric pattern $\Tilde{X}^{GRAPH}_t$, feature matrix $X_t$
   \STATE {\bfseries Output:} prediction $\hat{y}_{t+1:t+q}$
   \STATE $\Tilde{X}^{TOT}_t = \Tilde{X}^{TIME}_t + \Tilde{X}^{GRAPH}_t$ 
   \IF{Skip Layer True}
    \STATE $\Tilde{X}^{TOT}_t = \text{CONCAT}(\Tilde{X}^{TOT}_t,X_t )$
   \ENDIF
   \STATE $Z^{(0)}_t = \Tilde{X}^{TOT}_t$
   \FOR{$l=1$ {\bfseries to} $L^{''}$}
   \STATE $Z^{(l)}_t = \phi(\text{FC}^{(l)}(Z^{(l-1)}_t))$ 
   \ENDFOR
   \STATE $\hat{y}_{t+1:t+q} = Z^{(L^{''})}_t$ 
\end{algorithmic}
\end{algorithm}

The process begins by summing up the two patterns, and the result is denoted as $\Tilde{X}^{TOT}_t$. Then, if a skip layer is considered, $\Tilde{X}^{TOT}_t$ is concatenated with the feature matrix $X_t$. Finally, the hidden layer, denoted as $Z^{(0)}_t$ is initialized with $\Tilde{X}^{TOT}_t$. Before obtaining the prediction vector, denoted as $\hat{y}_{t+1:t+q}$, the $l$-th layer of a fully connected model is applied, with the number of layers denoted as $L^{''}$. 

\section{Experimental Evaluation}
\label{section_5}
In this section, we delineate the methodology employed for comparing the Time-Geometric model with the baseline model. We initiate by specifying the dataset and the pre-processing steps taken into account. Subsequently, we introduce the chosen evaluation metrics designed to facilitate the comparison. Prior to presenting the results, we expound on the process of determining the hyperparameters of the model utilized in the analysis. The computations were performed using the Nvidia GPU A$100$-SXM$4$-$40$GB.

\subsection{Dataset, Pre-processing, and Evaluation Metrics}
For our analysis, we focus on the constituents of the Standard \& Poor 100 (S\&P100), which exclusively includes the large-cap companies of the Standard \& Poor 500. We gather daily observations of Close, Open, High, Low, and Volume for each stock from Yahoo Finance covering the period from January 04, 2015, to June 04, 2023. After excluding stocks with missing values, we obtain a dataset comprising $90$ time series, each with $5$ features, and approximately $2,117$ observations. Before normalizing each time series to achieve a mean of zero and a variance of one using a rolling window approach, we partition our dataset into training, validation, and test sets using the $0.6-0.2-0.2$ rule.\\
To compare results across different models, we consider the following evaluation metrics: Root Mean Square Error (RMSE), Mean Absolute Percentage Error (MAPE), and Mean Absolute Scaled Error (MASE) \cite{hyndman2006another}.
A detailed exposition of the pre-processing steps and the evaluation metrics is provided in Appendix \ref{appendix:preprocessing} and \ref{appendix:preliminaries_metric}, respectively.

\subsection{Configuration of the Models}
To efficiently train our model, it is crucial to define the hyperparameters and the considered optimization strategy. The baseline models are the neural networks used in the Time Component, namely LSTM, RNN, GRU, BiLSTM, BiRNN, BiGRU, Transformers, TCN; these baseline models correspond to Equation (\ref{eq:g_function_definition}). The Time-Geometric model involves combining the baseline models with GCN, corresponding to Equation (\ref{eq:f_function_definition}). The hyperparameters to optimize encompass the batch size, sequence length ($m$), whether the dropout layer and skip connection are considered, and for both the Temporal and Geometric Components: the number of layers, $L$ and $L^{'}$ respectively, the number of neurons, the dropout rate, and the activation function, $\phi(\cdot)$ and $\rho(\cdot)$ respectively. Moreover, 
in the Geometric Component, we also need to determine the hidden dimension of the LSTM layer and the direction of the graph. In the Fully Connected Component, the sole hyperparameter is the activation function $\phi(\cdot)$, as the number of layers was set to four with $128, 64, 32, 16$ neurons respectively. The optimization of hyperparameters was performed using the Optuna Python package \cite{akiba2019optuna}, the Adam algorithm \cite{kingma2014adam} served as our optimization algorithm, and the Mean Squared Error (MSE) with the $l_2$ regularization term was used as loss function. Additionally, we specified the number of epochs as $1,000$ and incorporated early stopping with a patience of $20$ epochs to mitigate the risk of overfitting. To optimize the hyperparameters, we employed an incremental optimization strategy. Initially, we optimized the hyperparameters for the baseline models. Subsequently, for the Time-Geometric model, we used the optimized hyperparameters of the baseline model for the Time Component and optimized only the hyperparameters for the Geometric Component. 
For more details on the hyperparameter optimization and their values, refer to Appendix \ref{appendix:hyperparameter}.

\subsection{Numerical Results}
\begin{figure}[ht]
\vskip 0.1in
\begin{center}
\centerline{\includegraphics[width=\columnwidth]{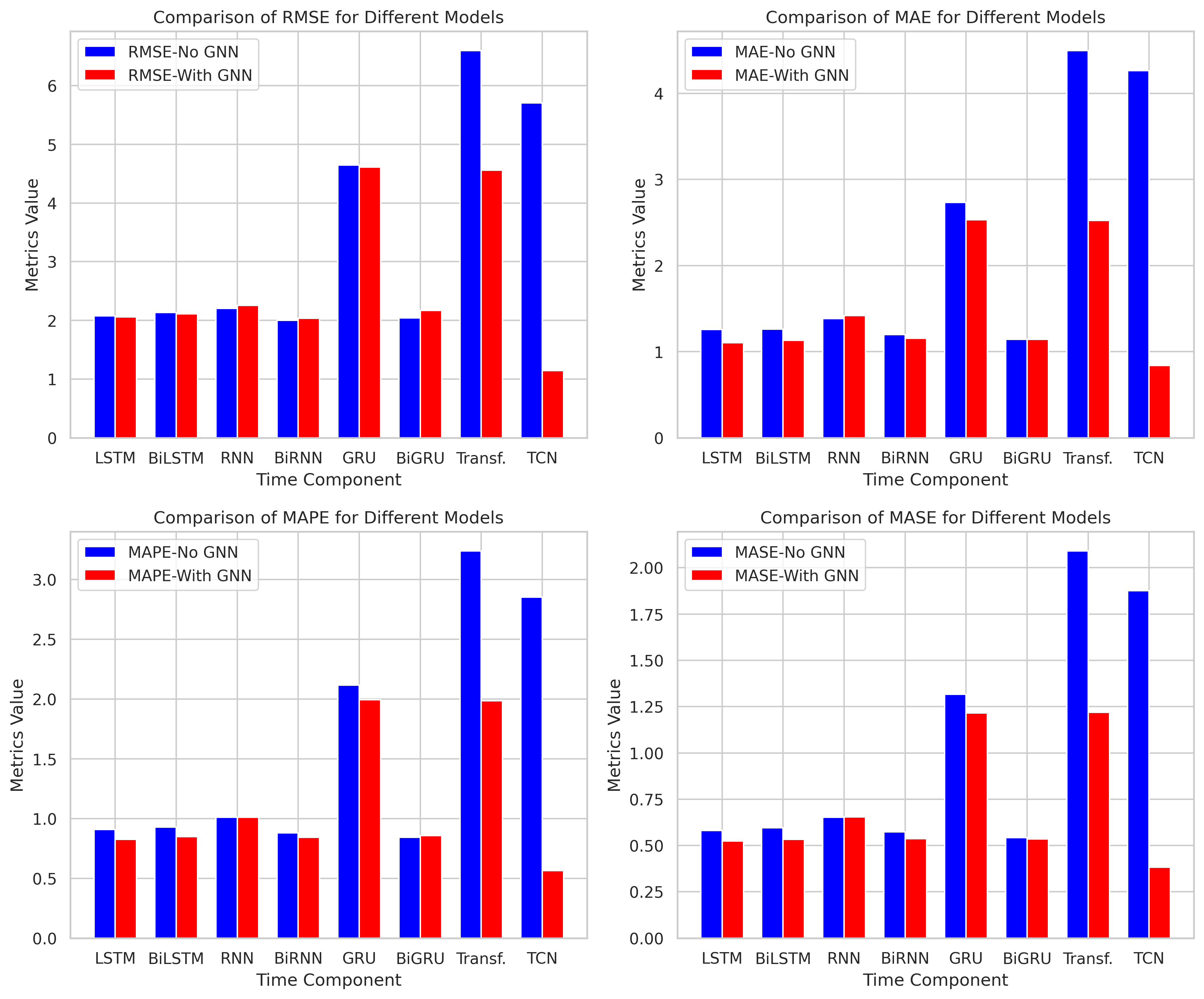}}
\caption{Average metric values for $1$ day prediction for all the models within the considered dataset are presented in the figure. The x-axis denotes the models used both as baselines and in the Time Component. The blue bar represents the average metric values for the respective baseline model, while the red bar signifies the average metric value for the Time-Geometric model.}
\label{fig:results_1day}
\end{center}
\vskip -0.1in
\end{figure}

We choose to forecast the `Close' price variable, denoted as $S^{*}$, and conduct predictions at various temporal intervals denoted as $q$: daily, weekly, and monthly, corresponding to $1$, $5$, and $20$ days. Furthermore, our feature matrix incorporates the variables `Close, High, Low, Open, and Volume' for each stock considered.\\
In Figure \ref{fig:results_1day} and Table \ref{tab:result_1day}, we present the comparison for the different evaluation metrics considered between our proposed model and the baseline model. Lower values indicate better model performance.

\begin{table}[h!]
\caption{The table displays the average metric values for a $1$ day prediction across all models in the considered dataset. In parentheses, we specify the Time Component considered in the Time-Geometric model, denoted in the table as TG. The best-performing model for each metric is highlighted in bold.}
\label{tab:result_1day}
\vskip 0.1in
\begin{center}
\begin{small}
\begin{sc}
\begin{tabular}{lccccc}
\toprule
Model       & RMSE & MAE & MAPE & MASE  \\
\midrule
LSTM       & 2.0743 & 1.2570 & 0.9074 & 0.5810 \\
BiLSTM     & 2.1349 & 1.2596 & 0.9276 & 0.5957 \\
RNN        & 2.2012 & 1.3841 & 1.0110 & 0.6520 \\
BiRNN     & 2.0006 & 1.1962 & 0.8797 & 0.5734 \\
GRU        & 4.6464 & 2.7335 & 2.1174 & 1.3166 \\
BiGRU      & 2.0438 & 1.1416 & 0.8441 & 0.5422 \\
Transf.    & 6.5977 & 4.4957 & 3.2390 & 2.0907 \\
TCN        & 5.7037 & 4.2612 & 2.8537 & 1.8767 \\
\hline
TG(LSTM)   & 2.0580 & 1.1044 & 0.8262 & 0.5248 \\
TG(BiLSTM) & 2.1141 & 1.1311 & 0.8487 & 0.5338 \\
TG(RNN)    & 2.2554 & 1.4192 & 1.0122 & 0.6553 \\
TG(BiRNN) & 2.0386 & 1.1529 & 0.8419 & 0.5377 \\
TG(GRU)    & 4.6131 & 2.5326 & 1.9950 & 1.2150 \\
TG(BiGRU)  & 2.1665 & 1.1420 & 0.8570 & 0.5360 \\
TG(Transf.)& 4.5617 & 2.5229 & 1.9869 & 1.2190 \\
TG(TCN)    & \textbf{1.1420} & \textbf{0.8376} & \textbf{0.5658} & \textbf{0.3828} \\
\bottomrule
\end{tabular}
\end{sc}
\end{small}
\end{center}
\vskip -0.1in
\end{table}
For ease of comparison, we compute the average for each considered metric across the dataset for each model. In Figure \ref{fig:results_1day}, the results for all the considered models are reported, both with the incorporation of GNN (our proposed framework) and without (baseline models). Notably, the impact of geometric patterns becomes evident only when the Time-Geometric model employs the Transformer and the RCN models for the Time Component, showcasing a noticeable improvement in average performance. In general, the use of a GNN slightly enhances the performance of the baseline model. This holds for several metrics. Such a trend persists when considering predictions at 5 and 20 days, as depicted in Figures \ref{fig:results_5day} and \ref{fig:results_20day}, along with Tables \ref{tab:result_5day} and \ref{tab:result_20day} in Appendix \ref{appendix:reults_metrics}.  

\section{Statistical Significance of the Inclusion of the Geometric Patterns} 
\label{section_6}
In this section, we aim to demonstrate the statistical significance of the inclusion of geometric patterns in enhancing time series forecasting. We initiate this analysis by applying various pairwise test statistics to the metric results. Subsequently, we conduct multiple comparisons to reinforce and validate the robustness of the pairwise comparisons. \\
Recalling the notation introduced in Section \ref{section_3}, let $K$ denote the number of algorithms compared, $N$ represents the number of datasets used, and $v_{u}^{j}$ indicate the evaluation metric of the $j$-th algorithm on the $u$-th dataset. In this research, we have compared $16$ algorithms across $90$ datasets, denoted by $K$ and $N$, respectively. For both types of tests, we set the significance level, denoted by $\alpha$, equal to $5\%$. This signifies our willingness to accept a $5\%$ risk of erroneously concluding that there is a significant effect or difference when, in reality, there is none (Type I error).  

\begin{table*}[t]
\caption{The table presents the outcomes of the tests, with Acceptance denoted as $A$ and Rejection as $R$. The test statistics for the Paired t-test, Wilcoxon test, and Sign test are encapsulated in brackets. The entry $\Delta$-\textit{``Model''} signifies the comparison between the baseline model and its corresponding Time-Geometric model. For the Paired t-test, we consider $89$ degrees of freedom, resulting in a critical value of $1.986$. In the Wilcoxon test, the critical value is $1.9599$, and for the Sign test, the critical value is $54.42$.}
\label{tab:pairwise_test_1day}
\vskip 0.1in
\begin{center}
\begin{small}
\begin{sc}
\setlength{\tabcolsep}{4.6pt}
\begin{tabular}{l|cccc|cccc|cccc}
\toprule
& \multicolumn{4}{c|}{paired t-test} & \multicolumn{4}{c|}{Wilcoxon test} & \multicolumn{4}{c}{Sign test}\\
Model & RMSE & MAE & MAPE & MASE & RMSE & MAE & MAPE & MASE & RMSE & MAE & MAPE & MASE  \\
\midrule
$\Delta$ LSTM       & A(0.4) & R(4.1) & R(4.6) & R(5.1) & A(1.2) & R(4.4) & R(4.5) & R(4.2) & A(50) & R(64) & R(64) & R(65) \\
$\Delta$ BiLSTM     & A(0.4) & R(4.0) & R(4.9) & R(6.7)& A(1.6) & R(5.9) & R(5.9) & R(5.6) & A(58) & R(73) & R(73) & R(73) \\ 
$\Delta$ RNN         & A(0.8) & A(0.5) & A(.03) & A(0.1)& A(0.2) & A(.03) & A(.04) & A(.05)  & A(46) & A(47) & A(47) & A(47) \\ 
$\Delta$ BiRNN      & A(1.0) & A(1.0) & A(1.9) & R(2.6) & A(0.7) & A(1.9) & R(2.3) & R(2.0) & A(40) & A(53) & A(53) & A(50) \\ 
$\Delta$ GRU         & A(1.7) & R(6.1) & R(8.9) & R(8.8)& A(1.3) & R(7.5) & R(7.5) & R(7.3) & A(47) & R(82) & R(82) & R(80) \\ 
$\Delta$ BiGRU       & R(2.7) & A(.01) & A(0.7) & A(0.6) & R(3.9) & R(2.3) & A(1.9) & A(1.3)& R(60) & A(54) & A(54) & A(53) \\ 
$\Delta$ Transf.     & R(4.8) & R(7.0) & R(12) & R(17)& R(8.2) & R(8.2) & R(8.2) & R(8.2) & R(90) & R(90) & R(90) & R(90) \\ 
$\Delta$ TCN         & R(7.8) & R(8.2) & R(13) & R(20) & R(8.2) & R(8.2) & R(8.2) & R(8.2) & R(90) & R(90) & R(90) & R(90) \\ 
\bottomrule
\end{tabular}
\end{sc}
\end{small}
\end{center}
\vskip -0.1in
\end{table*}

\subsection{Pairwise Comparison}
In the pairwise comparison, we systematically assess the evaluation metrics for each baseline model and its corresponding  Time-Geometric implementation (i.e., incorporating the Geometric Component) 
across multiple datasets. This involves scrutinizing the four metrics used to validate the improvement in the performance of the Time-Geometric model compared to its baseline model counterpart. \\
The null hypothesis for the pairwise comparison posits that the baseline and its Time-Geometric implementation perform equally well. Let $v_{u}^{j(BL)}$ denote the evaluation metric of the baseline model (denoted by $j(BL)$) on the $u$-th dataset, and $v_{u}^{j(TG)}$ denote the evaluation metric of the Time-Geometric inclusion (denoted by $j(TG)$) in the baseline model on the same dataset $u$. The null hypothesis is expressed as:
\begin{equation}
    \mathbf{H_0:} \qquad v_{u}^{j(BL)} = v_{u}^{j(TG)} \,.\label{eq:null_hypothesis_pairtest} 
\end{equation}
To analyze this null hypothesis, three different tests are considered: the Paired t-test, the Wilcoxon signed-ranks test \cite{wilcoxon1992individual}, and the Sign test \cite{salzberg1997comparing,sheskin2003handbook}.

\textbf{Paired t-test.} This test examines whether the average difference in the evaluation metric between $v_{u}^{j(BL)}$ and $v_{u}^{j(TG)}$ is significantly different from zero. Let $d_u = v_{u}^{j(BL)} - v_{u}^{j(TG)}$ be the difference for the $u$-th dataset between the $j(BL)$ and $j(TG)$ algorithms, and $\Bar{d}=\frac{1}{N}\sum_{u=1}^{N}d_u$ be the average difference. The null hypothesis, defined in Equation (\ref{eq:null_hypothesis_pairtest}), can be expressed as: ${\mathbf{H_{0}:} \mu_{d} = 0}$, where $\mu_{d}$ represent the true mean that is approximated with the sample mean, i.e., $\Bar{d}$. Finally, the Paired t-test statistic is computed as: $ z^{t} =\frac{\Bar{d}}{\sigma_{\Bar{d}}}$,
where $\sigma_{\Bar{d}}$ represents the standard deviation of the average difference, and $z^{t}$ follows a Student distribution with $N-1$ degrees of freedom.\\
\textbf{Wilcoxon signed-ranks test.} This non-parametric test ranks the differences $d_i$ for the two algorithms for each dataset $N$, ignoring the signs, and then compares the ranks for positive and negative differences. The Wilcoxon test is a more robust alternative to the Paired t-test, assuming commensurability of the differences. Let $B^{+}$ denote the sum of the ranks for which the $j(TG)$ algorithm performs better than the $j(BL)$ algorithm (i.e., $v_{u}^{j(TG)} < v_{u}^{j(BL)}$), and let $B^{-}$ represent the opposite case. The Wilcoxon statistic is then defined as:
\begin{equation*}
    z^{w} = \frac{B^{*} - \frac{1}{4}N(N+1)}{\sqrt{\frac{1}{24}N(N+1)(2N+1)}}\,,
\end{equation*}
where $B^{*} = \min \left( B^{+},B^{-}\right)$, and $z^{w} $ follows a normal distribution for large numbers of datasets.\\
\textbf{Sign test.} In the Sign test, we count the total number of times that the algorithm $j(TG)$ outperforms $j(BL)$, denoted by $|\#W|$ (i.e., $v_{u}^{j(TG)} < v_{u}^{j(BL)}$). Under the null hypothesis, each algorithm should outperform the other approximately $N/2$ times. We can rewrite Equation (\ref{eq:null_hypothesis_pairtest}) as $\mathbf{H_0: } p\{v_{u}^{j(TG)} < v_{u}^{j(BL)}\}=0.5$, with $p\{\cdot\}$ denoting probability. 
For $N>25$, $|\#W|$ is approximately distributed following a normal distribution with mean $\frac{N}{2}$ and variance $\frac{\sqrt{N}}{2}$. The difference in performance between algorithms $j(TG)$ and $j(BL)$ is statistically significant if:
$|\#W| \geq \frac{N}{2} + 1.96 \frac{\sqrt{N}}{2}$,
where $1.96$ is the critical value for a normal distribution at the $95\%$ confidence interval.\\
In Table \ref{tab:pairwise_test_1day}, we present the results of the test statistics for the evaluation metrics considered in the context of the $1$ day prediction. Entries marked with $A$ signify acceptance of the null hypothesis, while those marked with $R$ indicate rejection, and the corresponding statistical test values are enclosed in parentheses. Additionally, $\Delta$-\textit{``Model''} denotes the comparison between the baseline model and its counterpart incorporating GCNs. 
Results for the $5$ day and $20$ day predictions are detailed in Tables \ref{tab:pairwise_test_5day} and \ref{tab:pairwise_test_20day}, respectively, available in Appendix \ref{appendix:pairwise_test}.
Observing Table \ref{tab:pairwise_test_1day}, it becomes apparent that the statistical significance of the models is notably contingent on the choice of evaluation metrics. Specifically, only $\Delta$-Tranf., $\Delta$-TCN, and $\Delta$-RNN exhibit consistent statistical outcomes across all metrics. This suggests that the statistical significance of the inclusion of geometric patterns, and more broadly, the distinctions among the models, is influenced by the specific evaluation metrics employed. These observations persist for the $5$ day and $20$ day tests, as shown in Tables \ref{tab:pairwise_test_5day} and  \ref{tab:pairwise_test_20day}, respectively, in Appendix \ref{appendix:pairwise_test}.
In the case of the $5$ day test statistics, there is a more uniform trend in the results. This indicates that the prediction horizon also plays a crucial role in determining the statistical significance of a model relative to others.

\subsection{Multiple Comparison}
Relying solely on pairwise comparisons for each algorithm and its Time-Geometric implementation could potentially lead to the random rejection of the considered null hypothesis. To mitigate this, we also execute a multiple comparison, aimed at controlling the probability of making a Type I error in at least one individual comparison, commonly referred to as ``family-wise error'' \cite{demvsar2006statistical}. \\
In the multiple comparison, we analyze all the algorithms together for each dataset, considering one evaluation metric at a time. Specifically, the multiple comparison is executed using the Friedman test \cite{friedman1937use, friedman1940comparison}, where we rank the performance $v_{u}^{j}$ of each algorithm for each dataset. We sort the selected evaluation metric in increasing order, assigning a rank of $1$ to the first element, $2$ to the second, and so forth. Let $RK_{u}^{j}$ define the rank of the $j$-th algorithm on the $u$-th dataset for the selected metric. The Friedman test analyzes the average ranks of the algorithms, computed as 
$\Bar{RK}_j = \frac{1}{N}\sum_{u=1}^{N} {RK}_{u}^{j}$. 
The average ranks for each algorithm, each evaluation metric, and each forecasting horizon are reported in Table \ref{tab:average rank} in Appendix \ref{appendix:multiple_test}. The null hypothesis posits that all the algorithms are equivalent, implying their ranks are also equivalent. Thus, the Friedman test checks if the average ranks are statistically different from the mean among all the ranks. The Friedman statistic is computed as:
\begin{equation*}
    \chi_{F}^{2} = \frac{12 N}{K(K+1)}\left[\sum_{j=1}^{K}\Bar{RK}_j^{2} - \frac{K(K+1)^2}{4}\right]\,,
\end{equation*}
where $\chi_{F}^{2}$ follows a Chi-squared distribution with $K-1$ degrees of freedom when $N>10$ and $K>5$. The outcome of the Friedman test is reported in Table \ref{tab:fridman_test}. Observing Table \ref{tab:fridman_test}, the null hypothesis of the Friedman test is always rejected for each metric and each forecasting horizon considered, meaning that there exists a difference in the performance among the algorithms.
\begin{table}[t]
\caption{The table displays the Friedman test results for each metric considered, with Acceptance denoted as $A$ and Rejection as $R$, while the test statistics are reported in brackets. We consider $15$ degrees of freedom, resulting in a critical value of $27.48$.}
\label{tab:fridman_test}
\vskip 0.1in
\begin{center}
\begin{small}
\begin{sc}
\begin{tabular}{lccc}
\toprule
Metric       & $1$-Day & $5$-Day & $20$-Day   \\
\midrule
RMSE       & R(1041)  & R(1068) & R(654) \\
MAE     & R(1025)  & R(1034) & R(652)  \\
MAPE        &  R(1025)  & R(1034) & R(652)   \\
MASE     &  R(1025)  & R(1040) & R(653)   \\
\bottomrule
\end{tabular}
\end{sc}
\end{small}
\end{center}
\vskip -0.1in
\end{table}
In order to study such a difference, we can perform the Nemenyi test \cite{nemenyi1963distribution}, where we compare all the algorithms to each other. Specifically, we state that the difference in performance of two algorithms is statistically significant if their average rank difference is at least larger than a critical difference, denoted by 
$CD=q_{\alpha}\sqrt{\frac{K(K+1)}{6N}}$, 
where $q_{\alpha}$ is a critical value derived from the Standardized Range statistics and divided by $\sqrt{2}$ \cite{demvsar2006statistical}. 
Substituting the corresponding value into the equation for the critical value, we determine that $CD$ is set at $2.50$.
\begin{figure}[ht]
\vskip 0.1in
\begin{center}
\centerline{\includegraphics[width=\columnwidth]{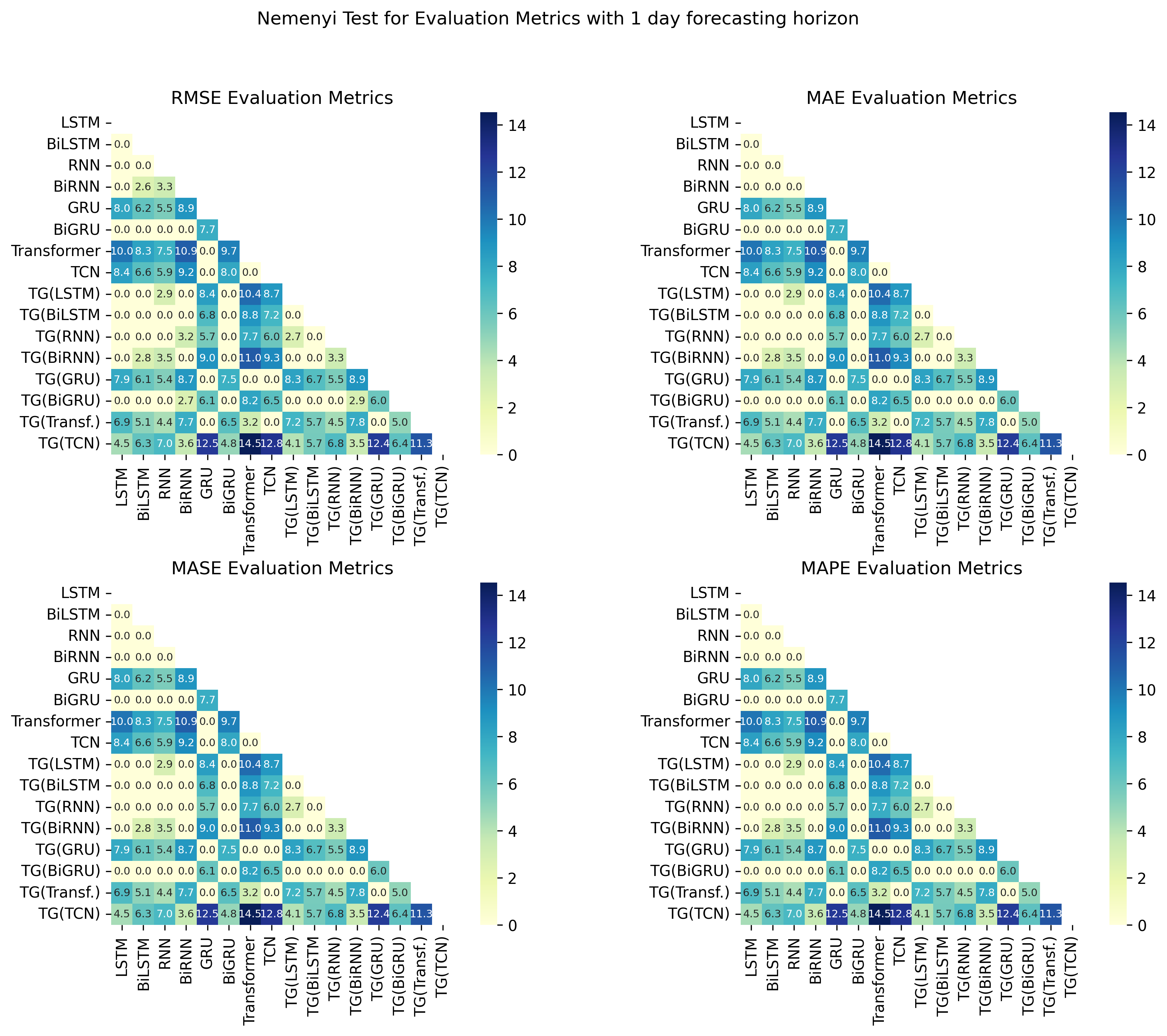}}
\caption{Namenyi test for $1$ day forecasting. Non-statistically significant relationships are highlighted in yellow.}
\label{fig:multiple_com_1day}
\end{center}
\vskip -0.1in
\end{figure}
Figure \ref{fig:multiple_com_1day} presents the results of the Nemenyi test for all the evaluation metrics considered in the context of a 1 day forecasting horizon. The comparative analysis for the 5 day and 20 day predictions is detailed in Figure \ref{fig:multiple_com_5day} and \ref{fig:multiple_com_20day} in Appendix \ref{appendix:multiple_test}, respectively. In all figures, non-statistically significant results are highlighted in yellow. Upon scrutinizing the ranks across models for each evaluation metric, we arrive at a conclusion consistent with the pairwise comparison. The choice of evaluation metrics and the forecasting horizon significantly influence the determination of a model's statistical significance. Notably, the Time-Geometric model underscores the statistical relevance of geometric patterns in enhancing the predictive accuracy of the baseline model.

\section{Conclusion}
\label{section_7}
In this study, we introduce the Time-Geometric model designed to leverage both temporal and geometric patterns for univariate financial time series forecasting. This is accomplished by integrating Graph Convolutional Networks into the baseline neural network model for time series, where the graph representation is derived from the visibility graph. We assess the performance of our model in comparison to various baseline models using a dataset comprising $90$ financial time series. Subsequently, we demonstrate that the statistical significance observed among the models is contingent on the choice of evaluation metric and forecasting horizon. Furthermore, we uncover the statistical significance of incorporating geometric patterns in enhancing forecasting accuracy.
Future research endeavors will concentrate on elucidating the behavior of evaluation metrics and exploring the statistical significance of the inclusion of geometric patterns across diverse time series datasets. 

\newpage
\bibliography{paper}

@article{lacasa2008time,
  title={From time series to complex networks: The visibility graph},
  author={Lacasa, Lucas and Luque, Bartolo and Ballesteros, Fernando and Luque, Jordi and Nuno, Juan Carlos},
  journal={Proceedings of the National Academy of Sciences},
  volume={105},
  number={13},
  pages={4972--4975},
  year={2008},
  publisher={National Academy of Sciences}
}

@misc{Bergillos2020visibility,
  title        = "Ts2vg: Time series to visibility graphs",
  author       = {Carlos Bergillos},
  howpublished = "\url{https://pypi.org/project/ts2vg/}",
  year         = 2020,
  note         = "Accessed: 2023-08-09"
}

@article{stephen2015visibility,
  title={Visibility Graph Based Time Series Analysis},
  author={Stephen, Mutua and Gu, Changgui and Yang, Huijie},
  journal={PloS One},
  volume={10},
  number={11},
  pages={e0143015},
  year={2015},
  publisher={Public Library of Science San Francisco, CA USA}
}

@book{Bishop2016,
  author={Bishop, Christopher M},
  title={Pattern Recognition and Machine Learning},
  year={2016},
  publisher={Springer New York, NY},
  edition={1}
}

@book{Tsay2005,
  title={Analysis of Financial Time Series},
  author={Tsay, Ruey S},
  year={2005},
  publisher={John Wiley \& Sons},
}

@book{lamberton2011introduction,
  title={Introduction to Stochastic Calculus Applied to Finance},
  author={Lamberton, Damien and Lapeyre, Bernard},
  year={2011},
  publisher={CRC press}
}

@book{Box2015,
  author={Box, George EP and Jenkins, Gwilym M and Reinsel, Gregory C and Ljung, Greta M},
  title={Time Series Analysis: Forecasting and Control},
  year={2015},
  publisher={John Wiley \& Sons},
  edition={5}
}

@article{Black1973pricing,
  title={The pricing of options and corporate liabilities},
  author={Black, Fischer and Scholes, Myron},
  journal={Journal of Political Economy},
  volume={81},
  number={3},
  pages={637--654},
  year={1973},
  publisher={The University of Chicago Press}
}

@book{tankov2003financial,
  title={Financial Modelling with Jump Processes},
  author={Tankov, Peter and Cont, Rama},
  year={2003},
  publisher={CRC Press}
}

@article{elman1990finding,
  title={Finding structure in time},
  author={Elman, Jeffrey L},
  journal={Cognitive Science},
  volume={14},
  number={2},
  pages={179--211},
  year={1990},
  publisher={Wiley Online Library}
}

@article{hua2019deep,
  title={Deep learning with long short-term memory for time series prediction},
  author={Hua, Yuxiu and Zhao, Zhifeng and Li, Rongpeng and Chen, Xianfu and Liu, Zhiming and Zhang, Honggang},
  journal={IEEE Communications Magazine},
  volume={57},
  number={6},
  pages={114--119},
  year={2019},
  publisher={IEEE}
}

@book{Goodfellow2016book,
    title={Deep Learning},
    author={Goodfellow, Ian and Bengio, Yoshua and Courville, Aaron },
    publisher={MIT Press},
    note={\url{http://www.deeplearningbook.org}},
    year={2016}
}

@article{vaswani2017attention,
  title={Attention is all you need},
  author={Vaswani, Ashish and Shazeer, Noam and Parmar, Niki and Uszkoreit, Jakob and Jones, Llion and Gomez, Aidan N and Kaiser, {\L}ukasz and Polosukhin, Illia},
  journal={Advances in Neural Information Processing Systems},
  volume={30},
  year={2017}
}

@misc{lea2016temporal,
      title={Temporal Convolutional Networks: A Unified Approach to Action Segmentation}, 
      author={Colin Lea and Rene Vidal and Austin Reiter and Gregory D. Hager},
      year={2016},
      eprint={1608.08242},
      archivePrefix={arXiv},
      primaryClass={cs.CV}
}

@book{mandelbrot2013fractals,
  title={Fractals and Scaling in Finance: Discontinuity, Concentration, Risk. Selecta volume E},
  author={Mandelbrot, Benoit B},
  year={2013},
  publisher={Springer Science \& Business Media}
}

@article{scarselli2008graph,
  title={The graph neural network model},
  author={Scarselli, Franco and Gori, Marco and Tsoi, Ah Chung and Hagenbuchner, Markus and Monfardini, Gabriele},
  journal={IEEE Transactions on Neural Networks},
  volume={20},
  number={1},
  pages={61--80},
  year={2008},
  publisher={IEEE}
}

@article{hamilton2017inductive,
  title={Inductive representation learning on large graphs},
  author={Hamilton, Will and Ying, Zhitao and Leskovec, Jure},
  journal={Advances in Neural Information Processing Systems},
  volume={30},
  year={2017}
}

@article{hochreiter1997long,
  title={Long short-term memory},
  author={Hochreiter, Sepp and Schmidhuber, J{\"u}rgen},
  journal={Neural Computation},
  volume={9},
  number={8},
  pages={1735--1780},
  year={1997},
  publisher={MIT press}
}

@article{Wu2021survey,
   title={A Comprehensive Survey on Graph Neural Networks},
   volume={32},
   ISSN={2162-2388},
   url={http://dx.doi.org/10.1109/TNNLS.2020.2978386},
   DOI={10.1109/tnnls.2020.2978386},
   number={1},
   journal={IEEE Transactions on Neural Networks and Learning Systems},
   publisher={Institute of Electrical and Electronics Engineers (IEEE)},
   author={Wu, Zonghan and Pan, Shirui and Chen, Fengwen and Long, Guodong and Zhang, Chengqi and Yu, Philip S.},
   year={2021},
   month=jan, pages={4–24} }

@inproceedings{Yu2018stgcn, 
   series={IJCAI-2018},
   title={Spatio-Temporal Graph Convolutional Networks: A Deep Learning Framework for Traffic Forecasting},
   url={http://dx.doi.org/10.24963/ijcai.2018/505},
   DOI={10.24963/ijcai.2018/505},
   booktitle={Proceedings of the Twenty-Seventh International Joint Conference on Artificial Intelligence},
   publisher={International Joint Conferences on Artificial Intelligence Organization},
   author={Yu, Bing and Yin, Haoteng and Zhu, Zhanxing},
   year={2018},
   month=jul, collection={IJCAI-2018} }

@article{cao2020spectral,
  title={Spectral temporal graph neural network for multivariate time-series forecasting},
  author={Cao, Defu and Wang, Yujing and Duan, Juanyong and Zhang, Ce and Zhu, Xia and Huang, Congrui and Tong, Yunhai and Xu, Bixiong and Bai, Jing and Tong, Jie and others},
  journal={Advances in Neural Information Processing Systems},
  volume={33},
  pages={17766--17778},
  year={2020}
}

@article{Zhao2020tgcn,
   title={{T-GCN}: A Temporal Graph Convolutional Network for Traffic Prediction},
   volume={21},
   ISSN={1558-0016},
   url={http://dx.doi.org/10.1109/TITS.2019.2935152},
   DOI={10.1109/tits.2019.2935152},
   number={9},
   journal={IEEE Transactions on Intelligent Transportation Systems},
   publisher={Institute of Electrical and Electronics Engineers (IEEE)},
   author={Zhao, Ling and Song, Yujiao and Zhang, Chao and Liu, Yu and Wang, Pu and Lin, Tao and Deng, Min and Li, Haifeng},
   year={2020},
   month=sep, pages={3848–3858} }

@inproceedings{xu2018graphwavelet,
title={Graph Wavelet Neural Network},
author={Bingbing Xu and Huawei Shen and Qi Cao and Yunqi Qiu and Xueqi Cheng},
booktitle={International Conference on Learning Representations},
year={2019},
url={https://openreview.net/forum?id=H1ewdiR5tQ},
}

@article{ruiz2020gatedrnn,
  title={Gated graph recurrent neural networks},
  author={Ruiz, Luana and Gama, Fernando and Ribeiro, Alejandro},
  journal={IEEE Transactions on Signal Processing},
  volume={68},
  pages={6303--6318},
  year={2020},
  publisher={IEEE}
}

@article{Cheng2018magnn,
title = {Financial time series forecasting with multi-modality graph neural network},
journal = {Pattern Recognition},
volume = {121},
pages = {108218},
year = {2022},
issn = {0031-3203},
doi = {https://doi.org/10.1016/j.patcog.2021.108218},
url = {https://www.sciencedirect.com/science/article/pii/S003132032100399X},
author = {Cheng, Dawei  and Yang, Fangzhou  and Xiang, Sheng  and Liu, Jin }
}

@article{Zhao2023wavenetatt,
title = {A new hybrid model for multi-step {WTI} futures price forecasting based on self-attention mechanism and spatial–temporal graph neural network},
journal = {Resources Policy},
volume = {85},
pages = {103956},
year = {2023},
issn = {0301-4207},
doi = {https://doi.org/10.1016/j.resourpol.2023.103956},
url = {https://www.sciencedirect.com/science/article/pii/S0301420723006670},
author = {Zhao, Geya  and Xue, Minggao  and Cheng, Li }
}

@inproceedings{Xiang2022thgnn,
author = {Xiang, Sheng and Cheng, Dawei and Shang, Chencheng and Zhang, Ying and Liang, Yuqi},
title = {Temporal and Heterogeneous Graph Neural Network for Financial Time Series Prediction},
year = {2022},
isbn = {9781450392365},
publisher = {Association for Computing Machinery},
address = {New York, NY, USA},
url = {https://doi.org/10.1145/3511808.3557089},
doi = {10.1145/3511808.3557089},
booktitle = {Proceedings of the 31st ACM International Conference on Information \& Knowledge Management},
pages = {3584–3593},
numpages = {10},
location = {Atlanta, GA, USA},
series = {CIKM '22}
}

@article{lazcano2023combined,
  title={A Combined Model Based on Recurrent Neural Networks and Graph Convolutional Networks for Financial Time Series Forecasting},
  author={Lazcano, Ana and Herrera, Pedro Javier and Monge, Manuel},
  journal={Mathematics},
  volume={11},
  number={1},
  pages={224},
  year={2023},
  publisher={MDPI}
}

@article{barrera2022rainfall,
  title={Rainfall prediction: A comparative analysis of modern machine learning algorithms for time-series forecasting},
  author={Barrera-Animas, Ari Yair and Oyedele, Lukumon O and Bilal, Muhammad and Akinosho, Taofeek Dolapo and Delgado, Juan Manuel Davila and Akanbi, Lukman Adewale},
  journal={Machine Learning with Applications},
  volume={7},
  pages={100204},
  year={2022},
  publisher={Elsevier}
}

@article{Shih2019ComparisonOT,
  title={Comparison of Time Series Methods and Machine Learning Algorithms for Forecasting Taiwan Blood Services Foundation's Blood Supply},
  author={Han M. Shih and Suchithra Rajendran},
  journal={Journal of Healthcare Engineering},
  year={2019},
  volume={2019},
  url={https://api.semanticscholar.org/CorpusID:203815441}
}

@article{Parmezan2019,
title = {Evaluation of statistical and machine learning models for time series prediction: Identifying the state-of-the-art and the best conditions for the use of each model},
journal = {Information Sciences},
volume = {484},
pages = {302-337},
year = {2019},
issn = {0020-0255},
doi = {https://doi.org/10.1016/j.ins.2019.01.076},
url = {https://www.sciencedirect.com/science/article/pii/S0020025519300945},
author = {Parmezan, Antonio Rafael Sabino and Souza, Vinicius M.A. and Batista, Gustavo E.A.P.A. }
}

@article{demvsar2006statistical,
  title={Statistical comparisons of classifiers over multiple data sets},
  author={Dem{\v{s}}ar, Janez},
  journal={The Journal of Machine Learning Research},
  volume={7},
  pages={1--30},
  year={2006},
  publisher={JMLR. org}
}

@article{chung2014empirical,
  title={Empirical evaluation of gated recurrent neural networks on sequence modeling},
  author={Chung, Junyoung and Gulcehre, Caglar and Cho, KyungHyun and Bengio, Yoshua},
  journal={arXiv preprint arXiv:1412.3555},
  year={2014}
}

@article{schuster1997bidirectional,
  title={Bidirectional recurrent neural networks},
  author={Schuster, Mike and Paliwal, Kuldip K},
  journal={IEEE Transactions on Signal Processing},
  volume={45},
  number={11},
  pages={2673--2681},
  year={1997},
  publisher={Ieee}
}

@inproceedings{graves2005bidirectional,
  title={Bidirectional {LSTM} networks for improved phoneme classification and recognition},
  author={Graves, Alex and Fern{\'a}ndez, Santiago and Schmidhuber, J{\"u}rgen},
  booktitle={International Conference on Artificial Neural Networks},
  pages={799--804},
  year={2005},
  organization={Springer}
}

@inproceedings{xiong2016dynamic,
  title={Dynamic memory networks for visual and textual question answering},
  author={Xiong, Caiming and Merity, Stephen and Socher, Richard},
  booktitle={International Conference on Machine Learning},
  pages={2397--2406},
  year={2016},
  organization={PMLR}
}

@article{kipf2016semi,
  title={Semi-supervised classification with graph convolutional networks},
  author={Kipf, Thomas N and Welling, Max},
  journal={arXiv preprint arXiv:1609.02907},
  year={2016}
}

@inproceedings{akiba2019optuna,
  title={Optuna: A next-generation hyperparameter optimization framework},
  author={Akiba, Takuya and Sano, Shotaro and Yanase, Toshihiko and Ohta, Takeru and Koyama, Masanori},
  booktitle={Proceedings of the 25th ACM SIGKDD International Conference on Knowledge Discovery \& Data Mining},
  pages={2623--2631},
  year={2019}
}

@article{hyndman2006another,
  title={Another look at measures of forecast accuracy},
  author={Hyndman, Rob J and Koehler, Anne B},
  journal={International Journal of Forecasting},
  volume={22},
  number={4},
  pages={679--688},
  year={2006},
  publisher={Elsevier}
}

@article{kingma2014adam,
  title={Adam: A method for stochastic optimization},
  author={Kingma, Diederik P and Ba, Jimmy},
  journal={arXiv preprint arXiv:1412.6980},
  year={2014}
}

@article{friedman1937use,
  title={The use of ranks to avoid the assumption of normality implicit in the analysis of variance},
  author={Friedman, Milton},
  journal={Journal of the American Statistical Association},
  volume={32},
  number={200},
  pages={675--701},
  year={1937},
  publisher={Taylor \& Francis}
}

@article{friedman1940comparison,
  title={A comparison of alternative tests of significance for the problem of m rankings},
  author={Friedman, Milton},
  journal={The Annals of Mathematical Statistics},
  volume={11},
  number={1},
  pages={86--92},
  year={1940},
  publisher={JSTOR}
}

@book{nemenyi1963distribution,
  title={Distribution-Free Multiple Comparisons},
  author={Nemenyi, Peter Bjorn},
  year={1963},
  publisher={Princeton University}
}

@article{salzberg1997comparing,
  title={On comparing classifiers: Pitfalls to avoid and a recommended approach},
  author={Salzberg, Steven L},
  journal={Data Mining and Knowledge Discovery},
  volume={1},
  pages={317--328},
  year={1997},
  publisher={Springer}
}

@book{sheskin2003handbook,
  title={Handbook of Parametric and Nonparametric Statistical Procedures},
  author={Sheskin, David J},
  year={2003},
  publisher={Chapman and hall/CRC}
}

@incollection{wilcoxon1992individual,
  title={Individual comparisons by ranking methods},
  author={Wilcoxon, Frank},
  booktitle={Breakthroughs in Statistics: Methodology and Distribution},
  pages={196--202},
  year={1992},
  publisher={Springer}
}

@article{benidis2022deep,
  title={Deep learning for time series forecasting: Tutorial and literature survey},
  author={Benidis, Konstantinos and Rangapuram, Syama Sundar and Flunkert, Valentin and Wang, Yuyang and Maddix, Danielle and Turkmen, Caner and Gasthaus, Jan and Bohlke-Schneider, Michael and Salinas, David and Stella, Lorenzo and others},
  journal={ACM Computing Surveys},
  volume={55},
  number={6},
  pages={1--36},
  year={2022},
  publisher={ACM New York, NY}
}

@article{mahmoud2021survey,
  title={A survey on deep learning for time-series forecasting},
  author={Mahmoud, Amal and Mohammed, Ammar},
  journal={Machine Learning and Big Data Analytics Paradigms: Analysis, Applications and Challenges},
  pages={365--392},
  year={2021},
  publisher={Springer}
}

@inproceedings{mahalakshmi2016survey,
  title={A survey on forecasting of time series data},
  author={Mahalakshmi, Ganapathy and Sridevi, S and Rajaram, Shyamsundar},
  booktitle={2016 International Conference on Computing Technologies and Intelligent Data Engineering (ICCTIDE'16)},
  pages={1--8},
  year={2016},
  organization={IEEE}
}

@book{brockwell1991time,
  title={Time Series: Theory and Methods},
  author={Brockwell, Peter J and Davis, Richard A},
  year={1991},
  publisher={Springer Science \& Business Media}
}

@article{rounaghi2016investigation,
  title={Investigation of market efficiency and financial stability between {S\&P} 500 and {L}ondon stock exchange: Monthly and yearly forecasting of time series stock returns using {ARMA} model},
  author={Rounaghi, Mohammad Mahdi and Zadeh, Farzaneh Nassir},
  journal={Physica A: Statistical Mechanics and its Applications},
  volume={456},
  pages={10--21},
  year={2016},
  publisher={Elsevier}
}

@book{hamilton2020time,
  title={Time Series Analysis},
  author={Hamilton, James D},
  year={2020},
  publisher={Princeton University Press}
}

@inproceedings{ariyo2014stock,
  title={Stock price prediction using the {ARIMA} model},
  author={Ariyo, Adebiyi A and Adewumi, Adewumi O and Ayo, Charles K},
  booktitle={2014 UKSim-AMSS 16th International Conference on Computer Modelling and Simulation},
  pages={106--112},
  year={2014},
  organization={IEEE}
}

@article{engle1982autoregressive,
  title={Autoregressive conditional heteroscedasticity with estimates of the variance of United Kingdom inflation},
  author={Engle, Robert F},
  journal={Econometrica: Journal of the Econometric Society},
  pages={987--1007},
  year={1982},
  publisher={JSTOR}
}

@article{lee2017comparative,
  title={Comparative study of volatility forecasting models: The case of {M}alaysia, {I}ndonesia, {H}ong {K}ong and {J}apan stock markets},
  author={Lee, San K and Nguyen, LT and Sy, Malick O},
  journal={Economics},
  volume={5},
  number={4},
  pages={299--310},
  year={2017}
}

@article{bollerslev1994arch,
  title={{ARCH} models},
  author={Bollerslev, Tim and Engle, Robert F and Nelson, Daniel B},
  journal={Handbook of Econometrics},
  volume={4},
  pages={2959--3038},
  year={1994},
  publisher={Elsevier}
}

@book{francq2019garch,
  title={{GARCH} models: Structure, Statistical Inference and Financial Applications},
  author={Francq, Christian and Zakoian, Jean-Michel},
  year={2019},
  publisher={John Wiley \& Sons}
}

@inproceedings{guo2016robust,
  title={Robust online time series prediction with recurrent neural networks},
  author={Guo, Tian and Xu, Zhao and Yao, Xin and Chen, Haifeng and Aberer, Karl and Funaya, Koichi},
  booktitle={2016 IEEE International Conference on Data Science and Advanced Analytics (DSAA)},
  pages={816--825},
  year={2016},
  organization={Ieee}
}

@inproceedings{ghosh2019stock,
  title={Stock price prediction using {LSTM} on {I}ndian Share Market},
  author={Ghosh, Achyut and Bose, Soumik and Maji, Giridhar and Debnath, Narayan and Sen, Soumya},
  booktitle={Proceedings of the 32nd International Conference on n Computer Applications in Industry and Engineering},
  volume={63},
  pages={101--110},
  year={2019}
}

@article{chen2023research,
  title={Research on Improved {GRU}-Based Stock Price Prediction Method},
  author={Chen, Chi and Xue, Lei and Xing, Wanqi},
  journal={Applied Sciences},
  volume={13},
  number={15},
  pages={8813},
  year={2023},
  publisher={MDPI}
}

@inproceedings{selvin2017stock,
  title={Stock price prediction using {LSTM}, {RNN} and {CNN}-sliding window model},
  author={Selvin, Sreelekshmy and Vinayakumar, R and Gopalakrishnan, EA and Menon, Vijay Krishna and Soman, KP},
  booktitle={2017 International Conference on Advances in Computing, Communications and Informatics (ICACCI)},
  pages={1643--1647},
  year={2017},
  organization={IEEE}
}

@article{roondiwala2017predicting,
  title={Predicting stock prices using {LSTM}},
  author={Roondiwala, Murtaza and Patel, Harshal and Varma, Shraddha},
  journal={International Journal of Science and Research (IJSR)},
  volume={6},
  number={4},
  pages={1754--1756},
  year={2017}
}

@inproceedings{zhang2017stock,
  title={Stock price prediction via discovering multi-frequency trading patterns},
  author={Zhang, Liheng and Aggarwal, Charu and Qi, Guo-Jun},
  booktitle={Proceedings of the 23rd ACM SIGKDD International Conference on Knowledge Discovery and Data Mining},
  pages={2141--2149},
  year={2017}
}

@inproceedings{ding2020hierarchical,
  title     = {Hierarchical Multi-Scale {G}aussian Transformer for Stock Movement Prediction},
  author    = {Ding, Qianggang and Wu, Sifan and Sun, Hao and Guo, Jiadong and Guo, Jian},
  booktitle = {Proceedings of the Twenty-Ninth International Joint Conference on Artificial Intelligence, {IJCAI-20}},
  publisher = {International Joint Conferences on Artificial Intelligence Organization},
  editor    = {Christian Bessiere},
  pages     = {4640--4646},
  year      = {2020},
  month     = {7},
  note      = {Special Track on AI in FinTech},
  doi       = {10.24963/ijcai.2020/640},
  url       = {https://doi.org/10.24963/ijcai.2020/640},
}

@article{lu2021cnn,
  title={A {CNN}-{BiLSTM}-{AM} method for stock price prediction},
  author={Lu, Wenjie and Li, Jiazheng and Wang, Jingyang and Qin, Lele},
  journal={Neural Computing and Applications},
  volume={33},
  pages={4741--4753},
  year={2021},
  publisher={Springer}
}

@article{lin2022new,
  title={A new attention-based {LSTM} model for closing stock price prediction},
  author={Lin, Yuyang and Huang, Qi and Zhong, Qiyin and Li, Muyang and Li, Yan and Ma, Fei},
  journal={International Journal of Financial Engineering},
  volume={9},
  number={03},
  pages={2250014},
  year={2022},
  publisher={World Scientific}
}

@article{barabasi2013network,
  title={Network science},
  author={Barab{\'a}si, Albert-L{\'a}szl{\'o}},
  journal={Philosophical Transactions of the Royal Society A: Mathematical, Physical and Engineering Sciences},
  volume={371},
  number={1987},
  pages={20120375},
  year={2013},
  publisher={The Royal Society Publishing}
}

@inproceedings{you2022roland,
  title={ROLAND: graph learning framework for dynamic graphs},
  author={You, Jiaxuan and Du, Tianyu and Leskovec, Jure},
  booktitle={Proceedings of the 28th ACM SIGKDD Conference on Knowledge Discovery and Data Mining},
  pages={2358--2366},
  year={2022}
}
\bibliographystyle{icml2024}

\newpage 
\appendix
\onecolumn
\section{Related work}
\label{appendix:related work}
Numerous studies have extensively explored diverse tasks related to time series, encompassing prediction and classification methodologies \cite{mahalakshmi2016survey, mahmoud2021survey, benidis2022deep}. This section presents a comprehensive overview of both deep learning and statistical models as applied to the forecasting of univariate financial time series.

\subsection{Time Series Forecasting with Statistical Models} 
Statistical models aim to discern and encapsulate distinct components within a time series, namely Trend, Seasonality, and Noise \cite{Box2015}. The AutoRegressive Moving Average (ARMA) model \cite{brockwell1991time} serves as a foundational statistical model for univariate time series analysis, amalgamating autoregressive and moving average elements. In a study by 
\citet{rounaghi2016investigation}, the ARMA model was employed for predicting stock returns in both the Standard and Poor's 500 (S\&P500) and the London Stock Exchange. However, limitations arise as the ARMA model assumes stationarity within the time series, presupposing constancy in mean and variance over time.
The AutoRegressive Integrated Moving Average (ARIMA) model \cite{hamilton2020time} extends the ARMA model by not explicitly assuming time series stationarity. Instead, the integrated component introduces differencing in the time series, achieving stationarity. 
\citet{ariyo2014stock} applied the ARIMA model to predict the New York Stock Exchange (NYSE) and the Nigeria Stock Exchange (NSE). Nevertheless, the ARIMA model reveals its constraints when faced with non-constant variance over time.
To address the challenge of evolving volatility, the AutoRegressive Conditional Heteroskedasticity (ARCH) model was introduced \cite{bollerslev1994arch}. In a study by 
\citet{engle1982autoregressive}, the ARCH model was employed to estimate mean and variance in the United Kingdom from 1958 to 1977. However, the ARCH model assumes homoscedastic volatility changes, signifying constant variance. Mitigating this limitation, the Generalized AutoRegressive Conditional Heteroskedasticity (GARCH) model was proposed \cite{francq2019garch}. 
\citet{lee2017comparative} conducted a comparative analysis between the ARIMA and GARCH models for predicting volatility in various stock exchanges.

\subsection{Time Series Forecasting with Deep Learning Models}
In contrast to statistical models, deep learning models offer the advantage of not requiring stationarity assumptions in time series data. Furthermore, they have demonstrated notable efficacy in capturing intricate non-linear patterns within such data. Among the class of neural network models tailored for time series analysis, Recurrent Neural Networks (RNNs) \cite{Goodfellow2016book} represent a fundamental model designed for sequential data processing. 
\citet{guo2016robust} applied RNNs for time series forecasting within the domain of anomaly detection. However, RNNs face challenges associated with the vanishing gradient problem, limiting their ability to capture long-term relationships within a time series. To address this limitation, the Long Short-Term Memory (LSTM) \cite{hochreiter1997long} model was introduced, along with its simplified counterpart, the Gated Recurrent Unit (GRU) \cite{chung2014empirical}.  
\citet{ghosh2019stock} utilized LSTM for predicting Indian Stock Exchanges, while  
\citet{chen2023research} compared the GRU model with other machine learning models for stock price prediction across various industries. Additionally, 
\citet{roondiwala2017predicting} employed LSTM to predict the NIFTY50 index. 
\citet{zhang2017stock} enhanced the LSTM model's performance by integrating the Discrete Fourier Transform into the memory cell. Furthermore, an attention mechanism was incorporated to augment the predictive capability of the LSTM model for various stock exchanges in \citet{lin2022new}. 
\citet{selvin2017stock} conducted a comparative analysis of the forecasting accuracy of LSTM, RNNs, and Convolutional Neural Networks (CNNs) on the Indian Stock Exchange. Finally, ensemble methods were proposed for predicting the Shanghai Composite index in \citet{ding2020hierarchical,lu2021cnn}.

\newpage
\section{Preliminaries}
This section offers a comprehensive discussion about the principles of Network Theory, the pre-processing stage, and the evaluation metrics considered in our study.

\subsection{Network Theory}
\label{appendix:preliminaries_network}
\begin{figure}[ht!]
\vskip 0.1in
\begin{center}
\centerline{\includegraphics[width=\textwidth]{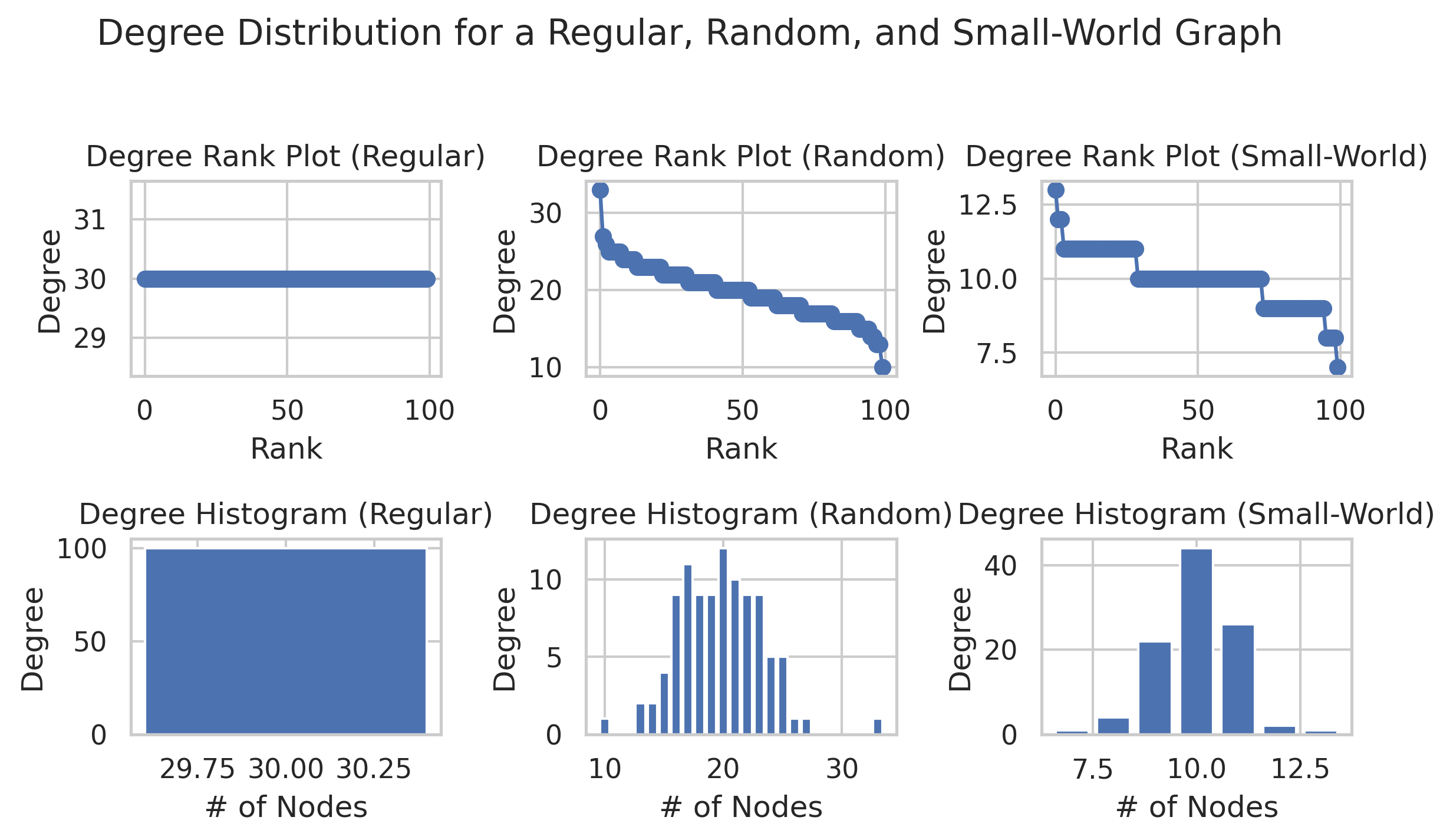}}
\caption{Example of degree distribution for a regular graph, a random graph, and a small-world graph.}
\label{fig:degree_dist}
\end{center}
\vskip -0.1in
\end{figure}

Let $G = (V, E)$ be a graph, where $V = \{v_1,\ldots,v_n\}$ is the
set of nodes, and $E \subseteq V \times V$ is the set of edges. The associated adjacency matrix $A$ for the graph $G$ is defined with entries $a_{uv}$ as follows:
\begin{equation*}
    a_{uv} = \begin{cases}
        1 \qquad \text{  if } (u, v) \in E\,, \\
        0 \qquad \text{ otherwise\,.}
    \end{cases}
\end{equation*}
This implies that nodes $u$ and $v$ are connected by an edge if and only if $a_{uv}=1$. Notably, the adjacency matrix $A$ is an $n \times n$ matrix. \\
Furthermore, a dynamic graph is conceptualized as a sequence of static graph snapshots indexed by discrete time. This approach, termed ``snapshot-based representation'' \cite{you2022roland}, defines a dynamic graph as $G = \{G_t\}^{T}_{t=1}$, where $G_t = (V_t, E_t)$ with $V_t \subseteq V$ and $E_t \subseteq E$. It is crucial to note that the set of nodes and edges may vary across different graph snapshot representations.\\
In this study, the visibility graph is employed to derive the graph representation of the time series. This algorithm is chosen for its capability to preserve the structural properties of time series data \cite{lacasa2008time}. Specifically, a periodic time series transforms into a regular graph, a random time series manifests as a random graph, and a fractal time series results in a small-world graph. The regular, random, and small-world graph properties 
are elucidated in \citet{barabasi2013network}, together with related topological characteristics. In a regular graph, each node $v$ has an identical number of connections, forming a highly ordered and structured network. This type of graph exhibits a regular and symmetrical pattern of connections, fostering uniformity throughout the network. Conversely, in a random graph, connections between nodes occur randomly, leading to a more unpredictable and less structured network. Here, $v$ and $u$ are connected with a probability $p$.  Lastly, the small-world graph is characterized by the presence of a few hubs: some nodes possess a large number of links, while many nodes have only a few links. This entails that most nodes can be reached from every other node in a small number of steps, while still maintaining local clusters or neighborhoods. To determine whether a graph is a regular, random, or small-world graph, the degree distribution of the graph must be analyzed. The degree of a node in a graph, denoted $k_u$, measures how connected or central that node is within the network and is defined as $k_u = \sum_{v=1}^{|V|}a_{uv}$.\\
Figure \ref{fig:degree_dist} exemplifies the characteristics of the degree distribution in a regular graph, a random graph, and a small-world graph. The total number of nodes is fixed at $100$, the degree for the regular graph is set to $30$, the probability of connecting two nodes in the random graph is set to $0.2$, and the small-world graph is configured with $10$ nearest neighbors to connect and a rewiring probability of $0.1$. 


\subsection{Pre-Processing}
\label{appendix:preprocessing}
In this research, we gather data comprising the ``Close, Open, High, Low'' prices, and the Volume for each stock considered within the Standard \& Poor 100 index. Normalization is employed to effectively capture non-linear patterns within the time series. To prevent potential data leakage during normalization, we adopt a rolling window approach. The dataset is normalized using the following method:
\begin{equation*}
    \Bar{x}^{i}_{t} = \frac{x^{i}_{t} - \mu^{i}_{t}}{\sigma^{i}_{t}}\,.
\end{equation*}
Here, $i \in \{\textit{Close, Open, High, Low, Volume}\}$, $x^{i}_{t}$ represents the $i$-th feature's value at time $t$, while $\mu^{i}_{t}$ and $\sigma^{i}_{t}$ denote the mean and standard deviation, respectively, of feature $i$. These statistical measures are computed using a rolling window, the length of which is a hyperparameter subject to optimization. 

\subsection{Evaluation Metrics}
\label{appendix:preliminaries_metric}
The purpose of employing evaluation metrics is to facilitate a comparative analysis of forecasting accuracy among the models utilized in this study. Four evaluation metrics were chosen for consideration: Root Mean Square Error (RMSE), Mean Absolute Error (MAE), Mean Absolute Percentage Error (MAPE), and Mean Absolute Scaled Error (MASE). For a dataset containing $M$ elements, these evaluation metrics are defined as follows:
\begin{align*}
    RMSE &= \sqrt{\frac{1}{M}\sum_{i=1}^{M}\left(y_i-\hat{y}_i\right)^2}\,, \\
    MAE &= \frac{1}{M}\sum_{i=1}^{M}|y_i-\hat{y}_i|\,, \\
    MAPE &= \frac{1}{M}\sum_{i=1}^{M}\left | \frac{y_i-\hat{y}_i}{y_i} \right |\,, \\
    MASE &= \frac{MAE}{\frac{1}{M-1}\sum_{i=2}^{M}|y_i- y_{i-1}|}\,, 
\end{align*}
where $y_i$ represents the true value for the $i$-th example, and $\hat{y}_i$ is the prediction of the model.  It is noteworthy that, for the evaluation metrics employed to gauge forecasting accuracy, smaller values indicate superior performance. Specifically, MASE was included in our selection due to its nature as a scale-free error metric. This implies that the MASE metric is less influenced by the scale of the data. Finally, the MASE metric compares the actual forecast against a naive forecast where the future one-step ahead value is set equal to the previous value. For the MASE metric, a value greater than $1$ suggests that the naive forecast outperforms the model's forecast, while a value lower than $1$ indicates the opposite.

\newpage
\section{Hyperparameters}
\label{appendix:hyperparameter}
This section outlines the hyperparameter optimization procedures implemented in the proposed analysis. Initially, we present the hyperparameters for the baseline models, followed by those for the Time-Geometric model.\\
For hyperparameter optimization, we employed the Python package Optuna \cite{akiba2019optuna}. Optuna facilitates optimization for a wide range of machine-learning models. Operating on the study-trial framework, Optuna defines a ``study'' as an optimization based on an objective function and a ``trial'' as a single execution of this optimization. The objective of a study is to identify the optimal set of hyperparameter values through multiple trials. We conducted $1000$ trials for each considered model. \\
Lastly, it is important to highlight that the choice of the loss function, regularization term, and optimization algorithm, as detailed in Section \ref{section_5}, has been made a priori for computational efficiency considerations.

\subsection{Baseline Models}
In our analysis, we include the following baseline models: Long Short-Term Memory (LSTM), Bidirectional LSTM (BiLSTM), Recurrent Neural Network (RNN), Bidirectional RNN (BiRNN), Gated Recurrent Unit (GRU), Bidirectional GRU (BiGRU), Transformer, and Temporal Convolutional Network (TCN). The hyperparameters for these baseline neural network models encompass learning rate, number of neurons, number of layers, sequence length, batch size, dropout rate, and activation function. Tables \ref{tab:Optimization_time_1day}, \ref{tab:Optimization_time_5day}, and \ref{tab:Optimization_time_20day} present the detailed hyperparameters for the baseline models corresponding to prediction horizons of $1$, $5$, and $20$ days.   

\begin{table}[h]
\caption{The table displays the hyperparameter values for a $1$ day prediction and the baseline models considered.}
\label{tab:Optimization_time_1day}
\vskip 0.1in
\begin{center}
\begin{small}
\begin{sc}
\setlength{\tabcolsep}{4.6pt}
\begin{tabular}{lcccccccc}
\toprule
Hyperparam.       & LSTM & BiLSTM & RNN & BiRNN & GRU & BiGRU & Tranf. & TCN  \\
\midrule
learning rate       & 0.000176 & 0.000258 & 0.000985 & 0.000656 & 0.000405 & 0.000134 & 0.000883 & 0.000935 \\
\# neurons      &  190 & 140 & 110 & 30 & 120 & 30 & - & 200 \\
\# layers      &  2 & 2 & 2 & 2 & 3 & 2 & 2 & - \\
sequence length      &  100 & 100 & 100 & 100 & 100 & 100 & 100 & 100 \\
batch size      &  40 & 160 & 160 & 80 & 160 & 80 & 40 & 160 \\
dropout rate      &  0.08 & 0.47 & 0.06 & 0.03 & 0.33 & 0.06 & 0.01 & 0.08 \\
Activation function      & selu & selu & selu & selu & elu & selu & elu & leaky relu  \\
\bottomrule
\end{tabular}
\end{sc}
\end{small}
\end{center}
\vskip -0.1in
\end{table}

\begin{table}[h]
\caption{The table displays the hyperparameter values for a $5$ day prediction and the baseline models considered.}
\label{tab:Optimization_time_5day}
\vskip 0.1in
\begin{center}
\begin{small}
\begin{sc}
\setlength{\tabcolsep}{4.6pt}
\begin{tabular}{lcccccccc}
\toprule
Hyperparam.       & LSTM & BiLSTM & RNN & BiRNN & GRU & BiGRU & Tranf. & TCN  \\
\midrule
learning rate       & 0.000731 & 0.000731 & 0.000168 & 5.254e-05 & 0.000165 & 0.000572 & 0.000884 & 8.053e-05 \\
\# neurons      &  190 & 120 & 140 & 190 & 50 & 180 & - & 50 \\
\# layers      &  4 & 6 & 2 & 5 & 4 & 2 & 2 & - \\
sequence length      &  100 & 100 & 100 & 100 & 100 & 100 & 100 & 100 \\
batch size      &  160 & 160 & 160 & 160 & 160 & 160 & 80 & 120 \\
dropout rate      &  0.29 & 0.37 & 0.32 & 0.23 & 0.36 & 0.23 & 0.27 & 0.41 \\
Activation function      & selu & elu & selu & selu & prelu & prelu & leaky relu & selu  \\
\bottomrule
\end{tabular}
\end{sc}
\end{small}
\end{center}
\vskip -0.1in
\end{table}

\begin{table}[h]
\caption{The table displays the hyperparameter values for a $20$ day prediction and the baseline models considered.}
\label{tab:Optimization_time_20day}
\vskip 0.1in
\begin{center}
\begin{small}
\begin{sc}
\setlength{\tabcolsep}{4.6pt}
\begin{tabular}{lcccccccc}
\toprule
Hyperparam.      & LSTM & BiLSTM & RNN & BiRNN & GRU & BiGRU & Tranf. & TCN  \\
\midrule
learning rate       & 0.000899 & 5.518e-05 & 0.000818 & 8.943e-05 & 0.000371 & 0.000117 & 0.000206 & 0.000178 \\
\# neurons      &  180 & 60 & 70 & 20 & 180 & 140 & - & 30 \\
\# layers      &  6 & 5 & 3 & 10 & 6 & 8 & 2 & - \\
sequence length      &  100 & 100 & 100 & 100 & 100 & 100 & 100 & 100 \\
batch size      &  80 & 160 & 80 & 160 & 160 & 80 & 160 & 160 \\
dropout rate      &  0.25 & 0.49 & 0.28 & 0.22 & 0.42 & 0.19 & 0.30 & 0.36 \\
Activation function      & selu & elu & elu & relu & selu & elu & leaky relu & elu  \\
\bottomrule
\end{tabular}
\end{sc}
\end{small}
\end{center}
\vskip -0.1in
\end{table}

\subsection{Time-Geometric Model}
The Time-Geometric model comprises three components, as elucidated in Section \ref{section_4}. For the Time Component, we adopt the same hyperparameters as those for the baseline models, as delineated in Tables \ref{tab:Optimization_time_1day}, \ref{tab:Optimization_time_5day}, and \ref{tab:Optimization_time_20day}, respectively, for predictions of $1$, $5$, and $20$ days. Concerning the Geometric Component, the hyperparameters encompass the dropout rate, the number of neurons and layers for the Graph Convolutional Network, the number of neurons for the LSTM layer, the presence of the dropout rate, and the specification of whether an undirected graph, denoted as ``None'',  or a directed graph, denoted as ``L-TO-R'' Left-to-Right), is considered to represent the time series. Conversely, the hyperparameters in the Fully Connected Component involve the consideration of the skip layer and the activation function, which, for simplicity, is set to match the activation function found in the baseline model. Tables \ref{tab:Optimization_1day}, \ref{tab:Optimization_5day}, and \ref{tab:Optimization_20day} outline the hyperparameters for the Time-Geometric model corresponding to prediction horizons of $1$, $5$, and $20$ days. 

\begin{table}[ht]
\caption{The table displays the hyperparameter values for a $1$ day prediction and the Time-Geometric model, denoted as TG, with the Time Component model values enclosed in brackets.}
\label{tab:Optimization_1day}
\vskip 0.1in
\begin{center}
\begin{small}
\begin{sc}
\setlength{\tabcolsep}{3pt}
\begin{tabular}{lcccccccc}
\toprule
Hyperparam.       & TG(LSTM) & TG(BiLSTM) & TG(RNN) & TG(BiRNN) & TG(GRU) & TG(BiGRU) & TG(Tranf.) & TG(TCN)  \\
\midrule
dropout rate     &  0.37 & 0.02 & 0.22 & 0.48 & 0.33 & 0.05 & 0.37 & 0.24 \\
\# neurons     & 40 & 190 & 110 & 90 & 160 & 120 & 110 & 60 \\
\# layer      &  9 & 9 & 7 & 7 & 8 & 10 & 2 & 7 \\
\# neurons lstm   & 40 & 20 & 100 & 30 & 180 & 30 & 180 & 80  \\
dropout?      &  True & False & True & True & True & False & True & True \\
Skip Layer      & True & True & False & True & True & True & True & True  \\
Direction      &  L-to-R & None & L-to-R & None & L-to-R & L-to-R & none & None\\
\bottomrule
\end{tabular}
\end{sc}
\end{small}
\end{center}
\vskip -0.1in
\end{table}

\begin{table}[ht]
\caption{The table displays the hyperparameter values for a $5$ day prediction and the Time-Geometric model, denoted as TG, with the Time Component model values enclosed in brackets.}
\label{tab:Optimization_5day}
\vskip 0.1in
\begin{center}
\begin{small}
\begin{sc}
\setlength{\tabcolsep}{3pt}
\begin{tabular}{lcccccccc}
\toprule
Hyperparam.       & TG(LSTM) & TG(BiLSTM) & TG(RNN) & TG(BiRNN) & TG(GRU) & TG(BiGRU) & TG(Tranf.) & TG(TCN)  \\
\midrule
dropout rate     &  0.25 & 0.17 & 0.13 & 0.4 & 0.06 & 0.03 & 0.03 & 0.01 \\
\# neurons     & 140 & 80 & 160 & 180 & 30 & 100 & 150 & 160 \\
\# layer      &  5 & 7 & 10 & 2 & 6 & 7 & 5 & 10 \\
\# neurons lstm   & 160 & 170 & 20 & 130 & 190 & 100 & 130 & 80  \\
dropout?      &  False & True & True & True & True & True & True & True \\
Skip Layer      & True & True & True & False & True & True & True & True  \\
Direction      &  L-to-R & None & L-to-R & None & None & L-to-R & none & None\\
\bottomrule
\end{tabular}
\end{sc}
\end{small}
\end{center}
\vskip -0.1in
\end{table}

\begin{table}[ht]
\caption{The table displays the hyperparameter values for a $20$ day prediction and the Time-Geometric model, denoted as TG, with the Time Component model values enclosed in brackets.}
\label{tab:Optimization_20day}
\vskip 0.1in
\begin{center}
\begin{small}
\begin{sc}
\setlength{\tabcolsep}{3pt}
\begin{tabular}{lcccccccc}
\toprule
Hyperparam.       & TG(LSTM) & TG(BiLSTM) & TG(RNN) & TG(BiRNN) & TG(GRU) & TG(BiGRU) & TG(Tranf.) & TG(TCN)  \\
\midrule
dropout rate     &  0.21 & 0.15 & 0.11 & 0.13 & 0.05 & 0.45 & 0.08 & 0.29 \\
\# neurons     & 130 & 160 & 110 & 160 & 190 & 20 & 90 & 180 \\
\# layer      &  5 & 2 & 6 & 5 & 9 & 2 & 2 & 3 \\
\# neurons lstm   & 50 & 20 & 190 & 170 & 80 & 20 & 190 & 50  \\
dropout?      &  False & False & True & False & True & False & True & False \\
Skip Layer      & False & True & False & True & False & False & False & False  \\
Direction      &  None & None & None & None & L-to-R & None & none & L-to-R\\
\bottomrule
\end{tabular}
\end{sc}
\end{small}
\end{center}
\vskip -0.1in
\end{table}

\newpage
\section{Results}
\label{appendix:results}
In this section, we present the results of the evaluation metrics for all the models considered in the analysis.

\subsection{Evaluation Metrics Results}
\label{appendix:reults_metrics}
Tables \ref{tab:result_5day} and \ref{tab:result_20day} present the results for all the models under consideration for the evaluation metrics, namely Root Mean Square Error (RMSE), Mean Absolute Error (MAE), Mean Absolute Percentage Error (MAPE), and Mean Absolute Scaled Error (MASE), for the $5$ and $20$ day prediction, respectively. It is important to note that a smaller value indicates better performance relative to the others. The best-performing model for each metric is highlighted in bold.
\begin{table}[ht]
\caption{The table displays the average metric values for a $5$ day prediction across all models in the considered dataset. In parentheses, we specify the Time Component considered in the Time-Geometric model, denoted in the table as TG. The best-performing model for each metric is highlighted in bold. }
\label{tab:result_5day}
\vskip 0.1in
\begin{center}
\begin{small}
\begin{sc}
\begin{tabular}{lcccc}
\toprule
Model       & RMSE & MAE & MAPE & MASE  \\
\midrule
LSTM       & 4.0979 & 3.0234 & 1.9594 & 1.3064 \\
BiLSTM     & 5.7802 & 4.2239 & 3.1294 & 2.0567 \\
RNN        & 3.1201 & 2.3216 & 1.5447 & 1.0180 \\
BiLSTM     & 3.0803 & 2.2928 & 1.5426 & 1.0238 \\
GRU        & 2.6417 & 2.0466 & 1.2799 & 0.8665 \\
BiGRU      & 3.0593 & 2.3167 & 1.4362 & 0.9515 \\
Transf.    & 9.4793 & 7.1196 & 4.8149 & 3.1075 \\
TCN        & 8.3543 & 6.2987 & 4.3513 & 2.8774 \\
\hline
TG(LSTM)   & 2.5442 & 1.7810 & 1.2149 & 0.7979 \\
TG(BiLSTM) & 4.4332 & 2.8562 & 2.1135 & 1.3274 \\
TG(RNN)    & 2.6913 & 1.8169 & 1.2785 & 0.8272 \\
TG(BiLSTM) & 3.1389 & 2.3015 & 1.5676 & 1.0375 \\
TG(GRU)    & \textbf{2.1209} & \textbf{1.6391} & \textbf{1.0352} & \textbf{0.6921} \\
TG(BiGRU)  & 2.7205 & 2.0417 & 1.2581 & 0.8224 \\
TG(Transf.)& 2.3148 & 1.7685 & 1.1219 & 0.7554 \\
TG(TCN)    & 3.5292 & 2.5441 & 1.8232 & 1.1920 \\
\bottomrule
\end{tabular}
\end{sc}
\end{small}
\end{center}
\vskip -0.1in
\end{table}

\begin{figure}[ht!]
\vskip 0.1in
\begin{center}
\centerline{\includegraphics[width=\textwidth]{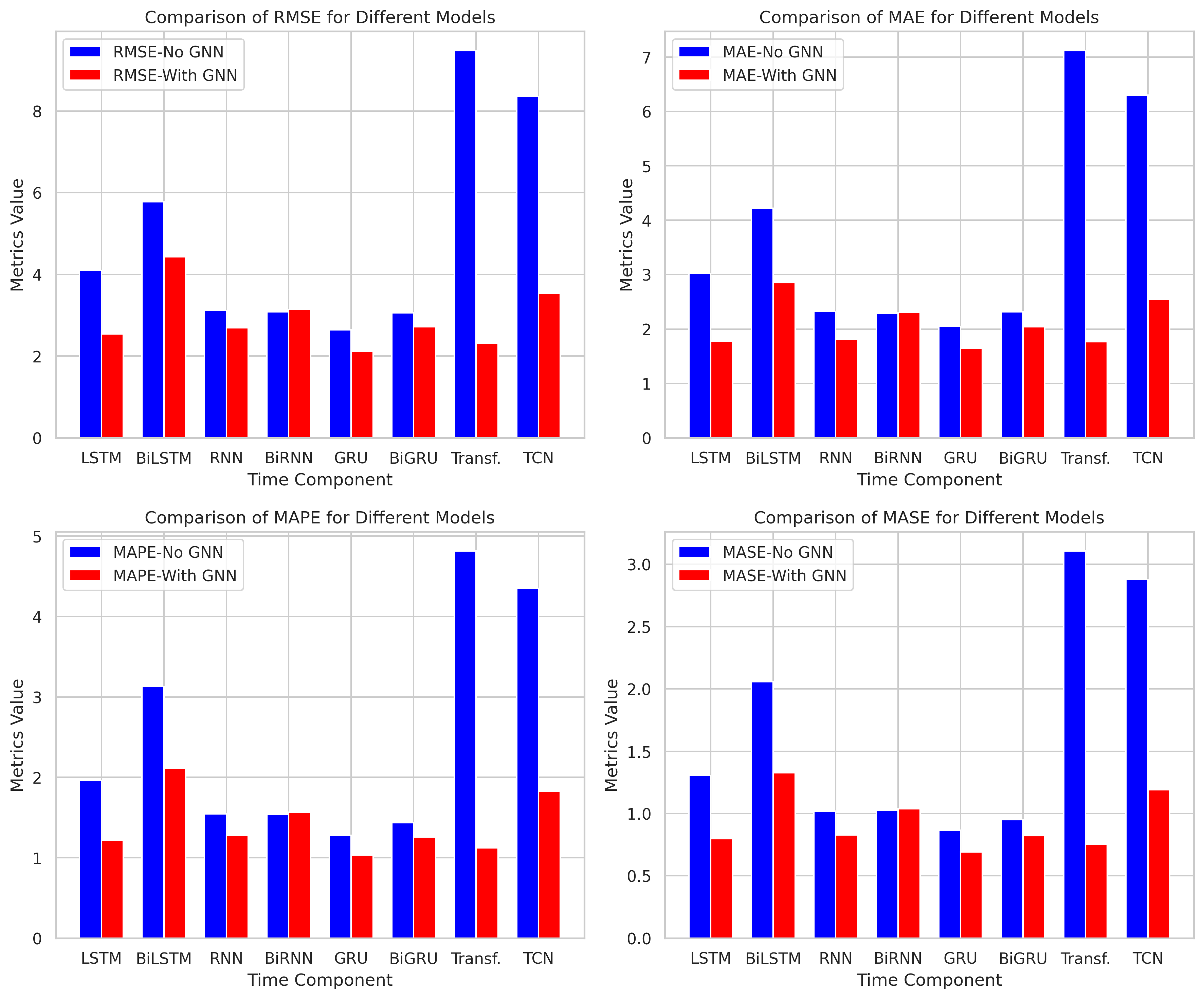}}
\caption{Average metric values for $5$ day prediction for all the models within the considered dataset are presented in the figure. The x-axis denotes the models used both as baselines and in the Time Component. The blue bar represents the average metric values for the respective baseline model, while the red bar signifies the average metric value for the Time-Geometric model.}
\label{fig:results_5day}
\end{center}
\vskip -0.1in
\end{figure}

\begin{figure}[ht!]
\vskip 0.1in
\begin{center}
\centerline{\includegraphics[width=\textwidth]{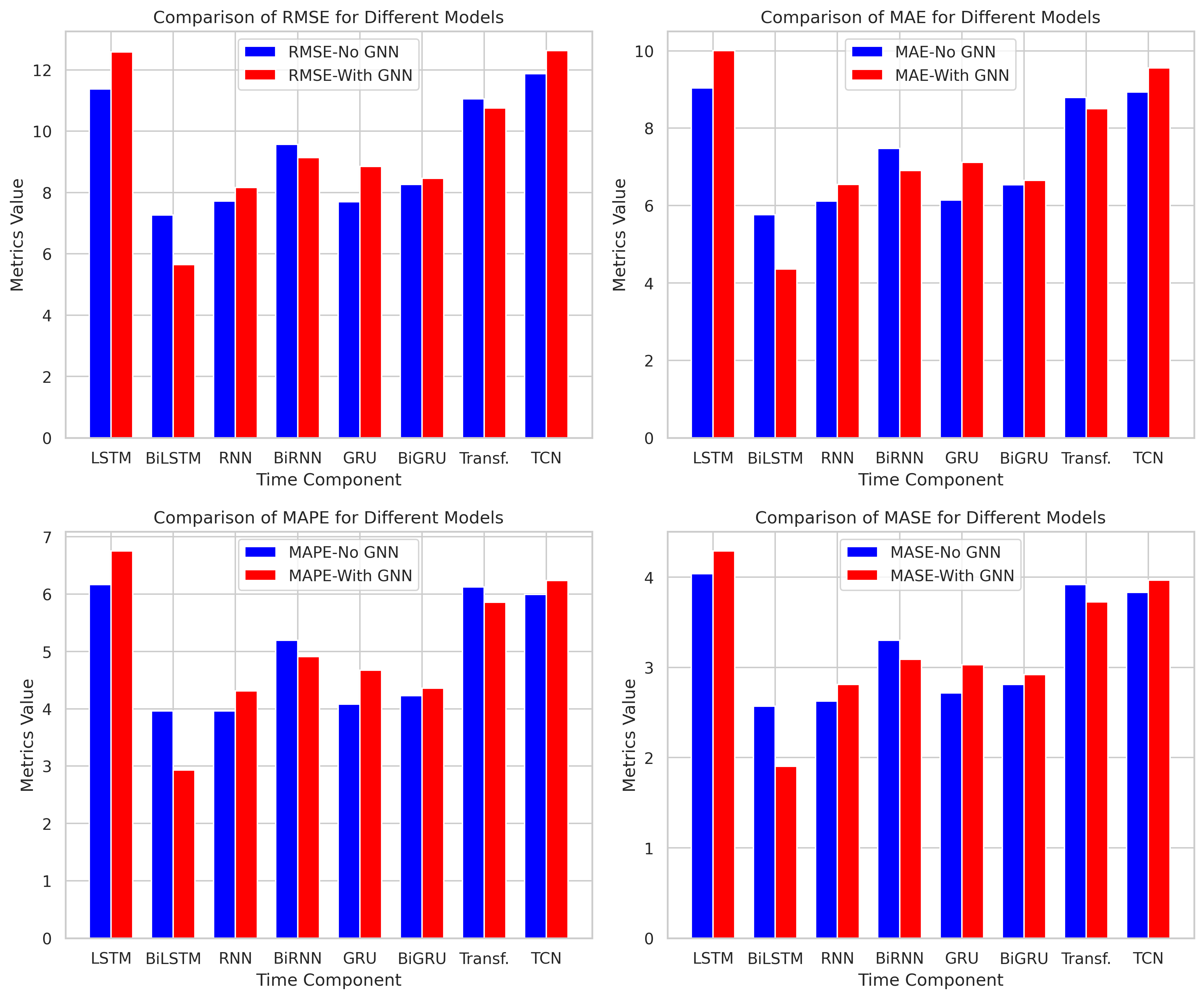}}
\caption{Average metric values for $20$ day prediction for all the models within the considered dataset are presented in the figure. The x-axis denotes the models used both as baselines and in the Time Component. The blue bar represents the average metric values for the respective baseline model, while the red bar signifies the average metric value for the Time-Geometric model.}
\label{fig:results_20day}
\end{center}
\vskip -0.1in
\end{figure}

\begin{table}[ht]
\caption{The table displays the average metric values for a $20$ day prediction across all models in the considered dataset. In parentheses, we specify the Time Component considered in the Time-Geometric model, denoted in the table as TG. The best-performing model for each metric is highlighted in bold.}
\label{tab:result_20day}
\vskip 0.1in
\begin{center}
\begin{small}
\begin{sc}
\begin{tabular}{lcccc}
\toprule
Model       & RMSE & MAE & MAPE & MASE  \\
\midrule
LSTM       & 11.3770 & 9.0394 & 6.1697 & 4.0384 \\
BiLSTM     & 7.2682 & 5.7614 & 3.9615 & 2.5693 \\
RNN        & 7.7205 & 6.1167 & 3.9599 & 2.6271 \\
BiLSTM     & 9.5645 & 7.4786 & 5.1974 & 3.3012 \\
GRU        & 7.6971 & 6.1448 & 4.0830 & 2.7182 \\
BiGRU      & 8.2659 & 6.5398 & 4.2292 & 2.8106 \\
Transf.    & 11.0544 & 8.7913 & 6.1235 & 3.9174 \\
TCN        & 11.8729 & 8.9351 & 5.9946 & 3.8302 \\
\hline
TG(LSTM)   & 12.5783 & 10.0061 & 6.7539 & 4.2900 \\
TG(BiLSTM) & \textbf{5.6507} & \textbf{4.3606} & \textbf{2.9326} & \textbf{1.9024} \\
TG(RNN)    & 8.1585 & 6.5460 & 4.3116 & 2.8110 \\
TG(BiLSTM) & 9.1312 & 6.9099 & 4.9102 & 3.0900 \\
TG(GRU)    & 8.8487 & 7.1136 & 4.6734 & 3.0290 \\
TG(BiGRU)  & 8.4649 & 6.6535 & 4.3596 & 2.9185 \\
TG(Transf.)& 10.7575 & 8.5082 & 5.8595 & 3.7253 \\
TG(TCN)    & 12.6266 & 9.5592 & 6.2406 & 3.9679 \\
\bottomrule
\end{tabular}
\end{sc}
\end{small}
\end{center}
\vskip -0.1in
\end{table}
In Figures \ref{fig:results_5day} and \ref{fig:results_20day}, we present histogram plots illustrating the distribution of evaluation metrics for all the models considered, focusing on the forecasting horizons of 5 days and 20 days, respectively. 

\newpage
\section{Statistical Significance}
In this section, we present the statistical significance analysis for all the models considered across various evaluation metrics. We start by detailing the outcomes of pairwise comparisons, followed by an exploration of multiple comparisons.

\subsection{Pairwise Comparison}
\label{appendix:pairwise_test}
Tables \ref{tab:pairwise_test_5day} and \ref{tab:pairwise_test_20day} display the results of the statistical tests conducted to assess potential statistical differences between the baseline models and their implementations including the Time-Geometric model for the $5$ and $20$ days prediction results, respectively.

\begin{table*}[ht]
\caption{The table presents the outcomes of the tests, with Acceptance denoted as $A$ and Rejection as $R$. The test statistics for the Paired t-test, Wilcoxon test, and Sign test are encapsulated in brackets. The entry $\Delta$-\textit{``Model''} signifies the comparison between the baseline model and its corresponding Time-Geometric model. For the Paired t-test, we consider $89$ degrees of freedom, resulting in a critical value of $1.986$. In the Wilcoxon test, the critical value is $1.9599$, and for the Sign test, the critical value is $54.42$.}
\label{tab:pairwise_test_5day}
\vskip 0.1in
\begin{center}
\begin{small}
\begin{sc}
\setlength{\tabcolsep}{4.6pt}
\begin{tabular}{l||cccc||cccc||cccc}
\toprule
& \multicolumn{4}{c||}{paired t-test} & \multicolumn{4}{c||}{Wilcoxon test} & \multicolumn{4}{c}{Sign test}\\
Model & RMSE & MAE & MAPE & MASE & RMSE & MAE & MAPE & MASE & RMSE & MAE & MAPE & MASE  \\
\midrule
$\Delta$ LSTM       & R(6.0) & R(6.6) & R(9.4) & R(11)& R(8.0) & R(8.1) & R(8.1) & R(8.0)   & R(87) & R(88) & R(88) & R(88) \\ 
$\Delta$ BiLSTM     & R(4.4) & R(5.6) & R(6.5) & R(6.4)& R(7.6) & R(7.9) & R(8.1) & R(8.1) & R(84) & R(86) & R(86) & R(86) \\ 
$\Delta$ RNN        & R(4.4) & R(5.6) & R(9.6) & R(8.1)& R(5.9) & R(7.3) & R(7.4) & R(7.1)& R(73) & R(80) & R(80) & R(80) \\ 
$\Delta$ BiRNN      & A(1.2) & A(0.2) & A(0.8) & A(1.1)& R(2.6) & A(0.6) & A(1.0) & A(1.2) & A(30) & A(44) & A(44) & A(43) \\ 
$\Delta$ GRU         & R(4.4) & R(4.4) & R(8.6) & R(5.9)& R(6.7) & R(6.4) & R(6.6) & R(6.4)& R(76) & R(76) & R(76) & R(75) \\
$\Delta$ BiGRU      & R(2.1) & R(2.2) & R(2.1) & A(1.9)& R(3.8) & R(3.6) & R(3.6) & R(3.5)& R(60) & R(58) & R(58) & R(57) \\
$\Delta$ Transf.    & R(7.1) & R(8.0) & R(18) & R(12) & R(8.2) & R(8.2) & R(8.2) & R(8.2)& R(90) & R(90) & R(90) & R(90) \\ 
$\Delta$ TCN        & R(6.1) & R(7.2) & R(16) & R(11) & R(8.2) & R(8.2) & R(8.2) & R(8.2)& R(90) & R(90) & R(90) & R(90) \\ 
\bottomrule
\end{tabular}
\end{sc}
\end{small}
\end{center}
\vskip -0.1in
\end{table*}

\begin{table*}[ht]
\caption{The table presents the outcomes of the tests, with Acceptance denoted as $A$ and Rejection as $R$. The test statistics for the Paired t-test, Wilcoxon test, and Sign test are encapsulated in brackets. The entry $\Delta$-\textit{``Model''} signifies the comparison between the baseline model and its corresponding Time-Geometric model. For the Paired t-test, we consider $89$ degrees of freedom, resulting in a critical value of $1.986$. In the Wilcoxon test, the critical value is $1.9599$, and for the Sign test, the critical value is $54.42$.}
\label{tab:pairwise_test_20day}
\vskip 0.1in
\begin{center}
\begin{small}
\begin{sc}
\setlength{\tabcolsep}{4.6pt}
\begin{tabular}{l|cccc|cccc|cccc}
\toprule
& \multicolumn{4}{c|}{paired t-test} & \multicolumn{4}{c|}{Wilcoxon test} & \multicolumn{4}{c}{Sign test}\\
Model & RMSE & MAE & MAPE & MASE & RMSE & MAE & MAPE & MASE & RMSE & MAE & MAPE & MASE  \\
\midrule
$\Delta$ LSTM       & A(1.4) & A(1.5) & A(1.5) & A(1.7)& A(1.0) & A(1.0) & A(0.9) & A(1.0)& A(42) & A(41) & A(41) & A(40) \\ 
$\Delta$ BiLSTM     & R(8.7) & R(8.8) & R(9.1) & R(7.5)& R(8.2) & R(8.2) & R(8.2) & R(8.2)& R(89) & R(90) & R(90) & R(89) \\ 
$\Delta$ RNN        & A(1.1) & A(1.5) & A(1.98) & R(2.1)& A(1.7) & R(2.1) & R(2.5) & R(2.5)& A(39) & A(39) & A(39) & A(39) \\ 
$\Delta$ BiRNN      & R(4.9) & R(5.7) & R(8.0) & R(8.3)& R(7.2) & R(7.3) & R(7.1) & R(7.2)& R(81) & R(82) & R(82) & R(80) \\ 
$\Delta$ GRU        & R(2.4) & R(2.5) & R(3.3) & R(3.1)& R(2.5) & R(2.5) & R(2.7) & R(2.6)& A(37) & A(35) & A(35) & A(34) \\ 
$\Delta$ BiGRU      & A(0.5) & A(0.4) & A(1.0) & A(0.9)& A(1.2) & A(0.8) & A(0.8) & A(0.7)& A(40) & A(43) & A(43) & A(39) \\ 
$\Delta$ Transf.    & A(1.1) & A(1.2) & R(3.1) & R(2.5)& R(4.0) & R(3.9) & R(3.8) & R(3.8) & R(66) & R(66) & R(66) & R(66) \\ 
$\Delta$ TCN        & A(1.1) & A(1.2) & A(1.2) & A(1.0)& A(1.5) & R(2.1) & A(1.8) & A(1.8)& A(40) & A(34) & A(34) & A(33) \\ 
\bottomrule
\end{tabular}
\end{sc}
\end{small}
\end{center}
\vskip -0.1in
\end{table*}

\subsection{Multiple Comparison}
\label{appendix:multiple_test}
Table \ref{tab:average rank} presents the average ranks for all the models across different evaluation metrics for the prediction horizons considered in the analysis — specifically, $1$, $5$, and $20$ days. The ranks are computed by comparing the evaluation metrics of all models for each dataset, assigning the value of $1$ to the best results, $2$ for the second-best, and so on. The average rank for each model is then computed. It is important to note that a lower rank suggests that the algorithm generally achieved better results compared to other models. Values in bold indicate the best rank for each metric.

\begin{table*}[ht]
\caption{The table presents the average ranks for the algorithms computed on each dataset. Values in bold indicate the best rank for each metric.}
\label{tab:average rank}
\vskip 0.1in
\begin{center}
\begin{small}
\begin{sc}
\setlength{\tabcolsep}{4pt}
\begin{tabular}{l|cccc|cccc|cccc}
\toprule
& \multicolumn{4}{c|}{$1$-Day} & \multicolumn{4}{c|}{$5$-Day} & \multicolumn{4}{c}{$20$-Day}\\
Model & RMSE & MAE & MAPE & MASE & RMSE & MAE & MAPE & MASE & RMSE & MAE & MAPE & MASE  \\
\midrule
LSTM       & 5.71 & 6.86 & 6.78 & 6.86 
           & 10.92 & 11.23 & 11.05 & 11.23 
           & 12.38 & 12.40 & 12.41 & 12.40 \\  
BiLSTM    & 7.50 & 8.12 & 8.17 & 8.12 
           & 13.95 & 13.94 & 13.96 & 13.94 
           & 5.10 & 5.10 & 5.20 & 5.10 \\  
RNN       & 8.21 & 8.78 & 8.67 & 8.78 
           & 8.52 & 8.50 & 8.48 & 8.50 
           & 5.54 & 5.66 & 5.55 & 5.66 \\  
BiRNN      & 4.86 & 6.46 & 6.24 & 6.46 
           & 8.63 & 8.93 & 8.93 & 8.93 
           & 10.02 & 10.05 & 10.15 & 10.05 \\  
GRU       & 13.74 & 13.90 & 13.90 & 13.90 
           & 5.51 & 6.22 & 6.05 & 6.22 
           & 5.87 & 5.93 & 5.86 & 5.93 \\  
BiGRU     & 6.05 & 5.90 & 5.92 & 5.90 
           & 5.13 & 5.55 & 5.52 & 5.55 
           & 6.70 & 6.82 & 6.67 & 6.82 \\  
Transf.    & 15.75 & 15.65 & 15.67 & 15.65 
           & 15.27 & 15.34 & 15.30 & 15.34 
           & 12.47 & 12.61 & 12.68 & 12.61 \\  
TCN        & 14.07 & 14.92 & 14.81 & 14.92 
           & 15.18 & 15.25 & 15.27 & 15.25 
           & 11.27 & 10.65 & 10.54 & 10.65 \\  
TG(LSTM)   & 5.34 & 4.57 & 4.56 & 4.57 
           & 4.90 & 4.43 & 4.44 & 4.43 
           & 12.90 & 13.11 & 13.08 & 13.11 \\  
TG(BiLSTM) & 6.91 & 5.07 & 5.24 & 5.07 
           & 12.11 & 11.21 & 11.42 & 11.21 
           & \textbf{1.50} & \textbf{1.35} & \textbf{1.38} & \textbf{1.35} \\  
TG(RNN)    & 8.07 & 8.34 & 8.37 & 8.34 
           & 6.23 & 5.05 & 5.20 & 5.05 
           & 6.42 & 6.64 & 6.62 & 6.64 \\  
TG(BiRNN)  & 4.74 & 5.28 & 5.32 & 5.28 
           & 9.15 & 8.77 & 8.86 & 8.77 
           & 8.36 & 7.94 & 8.22 & 7.94 \\  
TG(GRU)    & 13.60 & 12.46 & 12.57 & 12.46 
           & \textbf{2.18} & \textbf{2.51} & \textbf{2.44} & \textbf{2.51} 
           & 7.27 & 7.43 & 7.40 & 7.43 \\  
TG(BiGRU)  & 7.60 & 5.15 & 5.28 & 5.15 
           & 3.96 & 4.26 & 4.30 & 4.26 
           & 7.21 & 7.33 & 7.34 & 7.33 \\  
TG(Transf.) & 12.56 & 12.48 & 12.52 & 12.48 
           & 3.47 & 3.85 & 3.78 & 3.85 
           & 11.63 & 11.70 & 11.74 & 11.70 \\  
TG(TCN)    & \textbf{1.23} & \textbf{1.97} & \textbf{1.90} & \textbf{1.97} 
           & 10.82 & 10.90 & 10.93 & 10.90 
           & 11.30 & 11.23 & 11.08 & 11.23 \\  
\bottomrule
\end{tabular}
\end{sc}
\end{small}
\end{center}
\vskip -0.1in
\end{table*}

\begin{figure}[ht]
\vskip 0.1in
\begin{center}
\centerline{\includegraphics[width=\columnwidth]{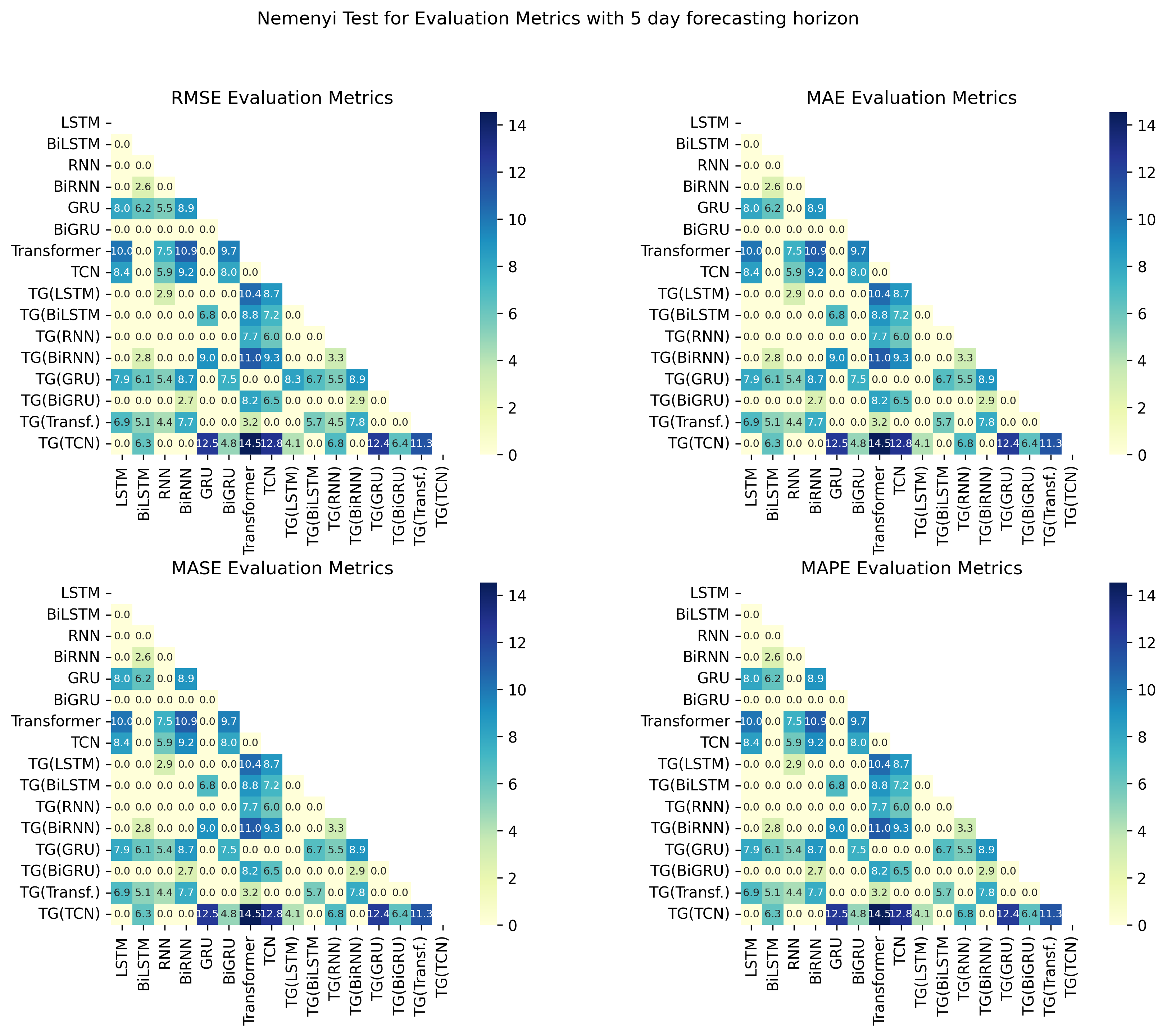}}
\caption{Namenyi test for $5$ day forecasting. Non-statistically significant relationship are highlighted in yellow.}
\label{fig:multiple_com_5day}
\end{center}
\vskip -0.1in
\end{figure}

\begin{figure}[ht]
\vskip 0.1in
\begin{center}
\centerline{\includegraphics[width=\columnwidth]{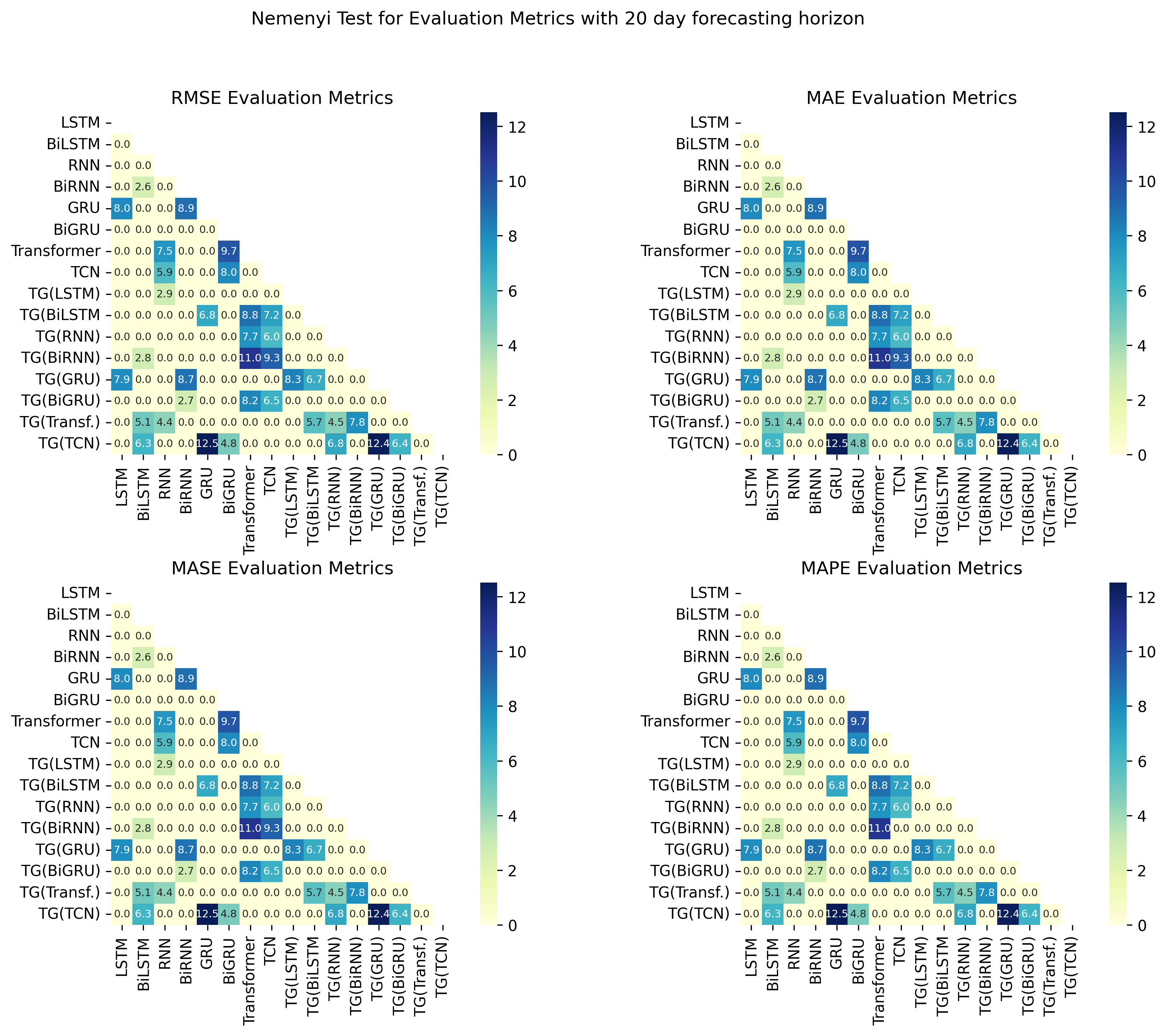}}
\caption{Namenyi test for $20$ day forecasting. Non-statistically significant relationship are highlighted in yellow.}
\label{fig:multiple_com_20day}
\end{center}
\vskip -0.1in
\end{figure}

In Figures \ref{fig:multiple_com_5day} and \ref{fig:multiple_com_20day}, the results of the Nemenyi test for all the evaluation metrics are presented for the $5$ and $20$ days forecasting horizons, respectively. This test assesses whether there is a statistically significant difference among the models utilized in the analysis. In the figures, yellow blocks indicate non-statistically significant differences between the models.

\end{document}